# BEYOND TRADITIONAL COATINGS, A REVIEW ON THERMAL SPRAYED FUNCTIONAL AND SMART COATINGS

D. Tejero-Martin[1], M. Rezvani Rad[2], A. McDonald[2], T. Hussain[1]*

[1]Faculty of Engineering, University of Nottingham, Nottingham, NG7 2RD, UK

[2]Faculty of Mechanical Engineering, University of Alberta, Edmonton, T6G 2G8, Canada

+44 115 951 3795, tanvir.hussain@nottingham.ac.uk

**Abstract**

Thermal spraying has been present for over a century, being greatly refined and optimised during this time. It has become nowadays a reliable and cost-efficient method to deposit thick coatings with a wide variety of feedstock materials and substrates. Thermal sprayed coatings have been successfully applied in fields such as aerospace or electricity production, becoming an essential component of today's industry. To overpass the traditional capabilities of those coatings, new functionalities and coherent responses are being integrated, opening the field of functional and smart coatings. The aim of this paper is to present a comprehensive review of the current state of functional and smart coatings produced using thermal spraying deposition. It will first describe the different thermal spraying technologies, with a focus on how different techniques achieve the thermal and kinetic energy required to form a coating. It will as well focus on the environment to which feedstock particles are exposed in terms of temperature and velocity. It will first deal with the state-of-the-art functional and smart coatings applied using thermal spraying techniques; a discussion will follow on the fundamentals on which the coatings are designed and the efficiency of its performance; finally, the successful applications, both current and potential will be described. The inherent designing flexibility of thermal sprayed functional and smart coatings has been exploited to explore exciting new possibilities on many different fields. Some applications include, but not limited to, prevention of bacteria contamination and infection on hygienic environments. Here, thermal spray has been used to efficiently deposit anti-microbial compounds on medical furniture and appliances and to develop biocidal and biocompatible coatings for prosthetic implants. The attachment of hard and soft foulers such as algae or molluscs, which represents a considerable issue for any marine or freshwater installation, can be prevented on components where the use of traditional antifouling strategies such as paints is not optimal such as polymers. Another interesting approach pursued is the development of superhydrophobic surfaces, with contact angles as high as 160° and slide angles below 5°, leading to high droplet mobility. This adds capabilities as self-cleaning or corrosion resistance in addition to the characteristic robustness of thermal sprayed coatings. The electric and magnetic properties of the feedstock materials have also led to the application of thermal spraying techniques in the creation of patterned structures with desired electromagnetic properties for their use on microelectronics. The possibility to intercalate layers of thermal sprayed materials doped with optical-reactive elements has led to the development of online and offline temperature sensors which can be readily integrated in current thermal barrier coatings. To finalise the examples of the many applications of thermal sprayed functional and smart coatings, autonomous self-healing or self-lubricant coatings have been developed. Advantage has been taken of a beneficial phase transformation triggered by the correspondent event (such as a crack or the tribological interactions respectively) to promote self-healing. Another approach has been the release of an encapsulated component which effectively heals the coating or provides lubrication when required. All these exciting developments pave the way for the numerous applications that are to come in the next decade, making the field of thermal sprayed coatings a unique opportunity for research and development.

1. Introduction

The selection of the materials to be used on industrial applications is dictated by their intrinsic properties, which must satisfy the specified needs for the component being designed and manufactured. A clear example would be structural components, where strength or fracture resistance is of the upmost importance. Nevertheless, any designed component will face a determined environment during service operation. This interaction can drastically limit its life-time or even change its properties to an extent where it will no longer satisfy the expected demands. Two common examples which illustrate these situations would be components in corrosive environments, such as offshore structures where the combination of salt and humidity corrodes the surface of the component, gradually weakening the structure (Ref 1); and high temperature applications such as turbine blades, where the elevated temperatures would damage unprotected components (Ref 2).

As the examples above illustrate, the interaction between the environment and the material plays a critical role, and at the heart of that interaction, lies the surface. The surface of any component represents the interface between the environment and the material, and any interaction that is due to take place will happen at the surface. In order to protect the surface, one successful strategy has been to deposit a relatively thin layer (when compared to the dimensions of the bulk material) of a different material, with the properties required to face the expected environment.

The development of coatings has pushed further away the inherent limits of materials, broadening the design possibilities on many cutting-edge fields. The advantages of combining the bulk properties of a substrate with the tailored capabilities of a layer at the surface opened a world of new opportunities. Nowadays, coatings are present in almost all of the most demanding environments and specialized applications. Following the examples above, offshore structures are coated with corrosion resistant coatings such as zinc and zinc-aluminium (Ref 3), and turbine blades are commonly coated with thermal barrier coatings such as yttria-stabilised zirconia (YSZ), with low thermal conductivity and good mechanical properties.

One of the aspects that has made coatings such a popular solution is the fact that a wide range of deposition techniques are available. To cover the whole catalogue of deposition techniques available is outside the scope of this review, which will focus on thermal spray technologies. Thermal spray comprises those deposition processes in which an energy source is used to heat the initial feedstock particles (which could be presented in the form of suspension, powder, wire or rod), being then accelerated and propelled towards the substrate using a gas stream (Ref 4).

The combined thermal and kinetic energies of the particles allows the bonding with the surface of the substrate upon impact, effectively building up the coating as the particles reach the surface. Another aspect that has contributed to the wide-spread use of thermal spraying is the flexibility in the choice of materials that can be deposited with these techniques. As a general definition, any material with the capability of melting without experiencing decomposition is suitable for thermal spraying (Ref 5).

Due to the unique combination of a wide range of deposition techniques and materials available for coatings, thermal spraying has been successfully applied in numerous fields, such as corrosion and oxidation resistance (Ref 6), high temperature protection (Ref 7), wear and erosion (Ref 8), abradable coatings and dimensional repairs (Ref 9,10), biomedical applications (Ref 11) and electronics (Ref 12). As it can be seen, the field of thermal spraying has represented a prolific match for coatings, allowing the production of highly capable systems. Such coating also present a great acceptance in the industry, excellent large-scale adaptability and suitable cost/efficiency, having undoubtedly proved its value. Despite the clear success and high rate of applicability of thermal sprayed coatings, there is an always increasing demand for systems capable of facing more aggressive environments, reliably performing at even higher temperatures, serving during longer periods of time or providing new, desireable functionalities, to name a few of the driving criteria. To overcome some of the current needs, new concepts within thermal sprayed coatings have been developed, being functional and smart the subject of investigation of this work.

One of the most successful routes in the development of thermal sprayed coatings has been the combination of a proven material system which, and due to the flexibility of allowed sprayed materials, is combined with an added component responsible for the novel functionality. The presence of a solid base of thoroughly investigated and field-tested thermal sprayed coatings has provided an unparalleled starting point for the development of more capable and tailor designed functional coatings. On its simplest definition, a functional coating can be described as a coating with an added functionality beyond the traditional protective capabilities (Ref 13). The classical protective case would be the already mentioned corrosion or wear protection. However, although these new functionalities provide functional coatings with a wide range of applications and possibilities, their behaviour is still passive on its interaction with the environment. A smart coating, on the other hand, aims to provide coatings with an active response to certain stimuli, generated either by intrinsic or extrinsic events (Ref 14,15). In summary, all smart coatings can be considered functional coatings due to the presence of a functionality beyond simple protection, but not all functional coatings can be categorised as smart due to the lack of an active response to external stimuli. It should be noted that the categorisation used in this review regarding the functional and smart coatings is not intended to be definitive— different definitions are present in existing literature. The distinction was chosen to provide a more structured approach to the review.

Several decades of investigation on the science behind thermal sprayed coatings and the relatively new addition of functional and smart coatings has provided an invaluable opportunity for numerous industrial applications. However further research is still required to provide cost-efficient methods with proven added value to the companies. With the unparalleled success of deposition techniques such as plasma spraying or high velocity oxy-fuel (HVOF) thermal spraying as an example, other thermal spray technologies still need to reach that level of market penetration. This will only be attainable through strong beneficial cases with a clear understanding of the processes involved. The addition of new capabilities through the introduction of functional and smart coatings represents an added opportunity for exciting, ground-breaking research. It is essential that the industry becomes involved too, setting the requirements and needs of a market more demanding than ever, in which these technologies can present a benefit.

In this work, and due to the importance that the different deposition techniques have on the produced functional and smart coatings, an overview of the main thermal spray technologies available is first presented. Then, attending to the division previously defined between functional and smart coatings, an extensive and comprehensive study of the current developments in the field is undertaken. With the use of thermal spraying techniques as common factor, the current state-of-the-art for functional and smart coatings is presented attending to the different functionalities achieved.

## 2. Thermal Spraying Technologies

Thermal spraying processes incorporate those technologies on which metallic or non-metallic coatings are deposited through the same principle. A heat source melts the feedstock material and a jet is used to impart kinetic energy to the molten particles. They then impinge the substrate surface and rapidly cool down to form a solid splat, continuously building up the desired thickness (Ref 16). A basic schematic of the thermal spray process can be seen in Figure 1. The flexibility on thermal sources and jet configurations give rise to a plethora of different deposition technologies, as presented in Figure 2, each one producing coatings with different microstructures and physical properties.

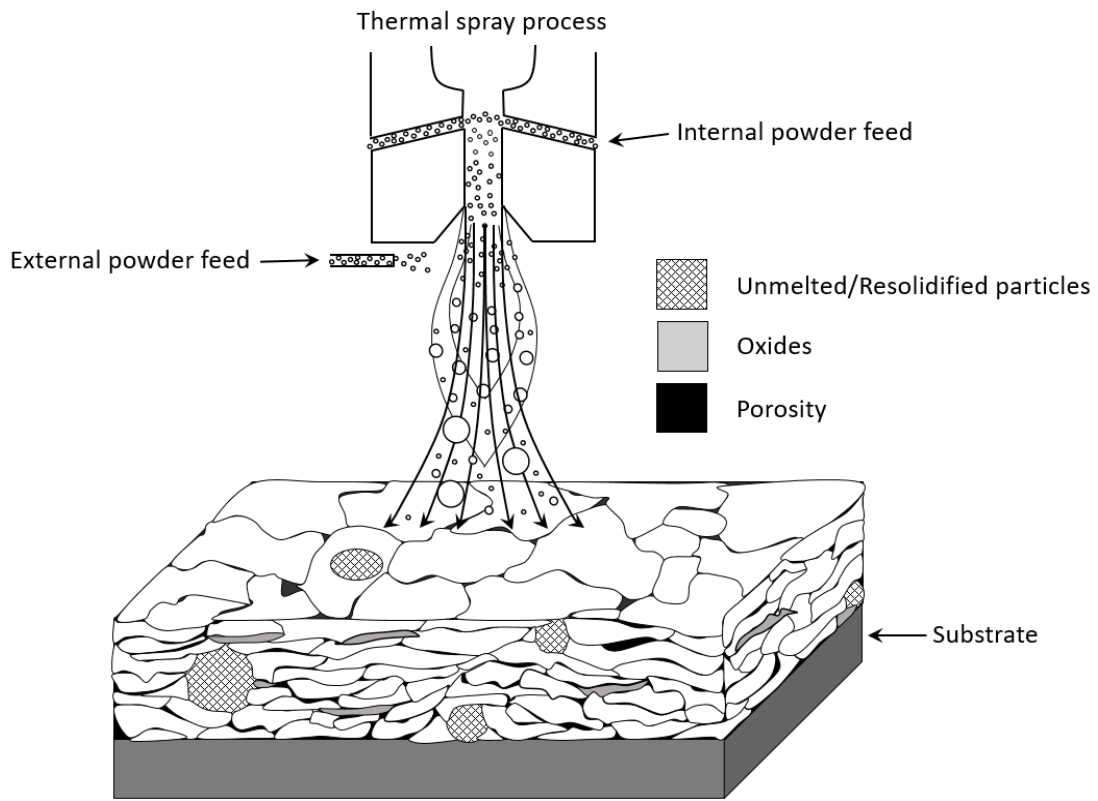

**Figure 1:** *Schematic representation of a thermal spraying process with the two main components, a heat source and a jet, and the main features of the produced coatings. Redrawn from* (Ref 4).

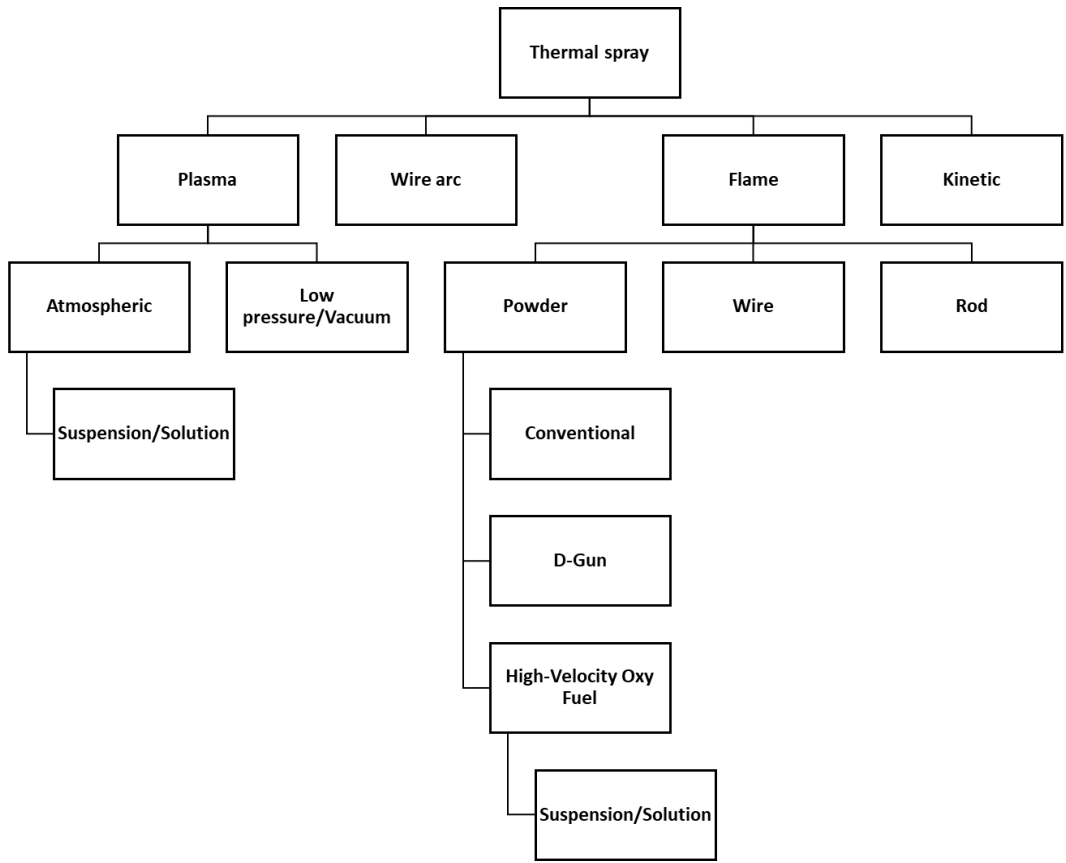

**Figure 2:** *Classification of the thermal spray family of deposition techniques. Redrawn from* (Ref 17).

The development of thermal spraying processes has been central to the evolution of functional and smart coatings, allowing new materials to be deposited and new substrates to be coated, broadening the range of accessible possibilities. This section presents an overview of the most common thermal spraying processes used in the fabrication of functional and smart coatings, with a brief description of their working principles.

2.1. Plasma spraying

Atmospheric plasma spraying (APS) uses thermal plasmas produced through direct current (DC) arc or radio frequency (RF) discharge as the heat source of the deposition process. This allows flame temperature over 8000 K (reaching as high as 14000 K in the jet core (Ref 16,18)) and particle velocities ranging from ~20 m s$^{-1}$ up to ~500 m s$^{-1}$ depending on the particle size distribution (Ref 19). The elevated temperatures produce a high proportion of particles melted, which in addition to the relatively high velocities, give rise to excellent deposition densities, bond strengths and low porosity coatings when compared to most thermal spraying processes (Ref 16). The high cost-efficiency and good quality of the coatings obtained by using APS has led to a successful implementation in numerous industries.

### 2.1.1. Suspension/solution precursor plasma spraying

Due to the need of adequate flowability for the feedstock powder, APS is limited to the deposition of particles with an approximate lower limit size of 10 – 100 µm (Ref 20,21). In order to allow the use of nano-scaled powders, different solutions have been developed as an alternative to the traditional injection of powder. The main representatives of these alternatives are suspension plasma spraying (SPS) and solution precursor plasma spraying (SPPS) (Ref 20–30). The differentiation factor between the two methods is shown in Figure 3, with the precipitation of the deposited particles in-flight in the case of SPPS as opposed to the direct deposition (apart from the physical changes related to the exposure to the high temperature in the flame) in SPS.

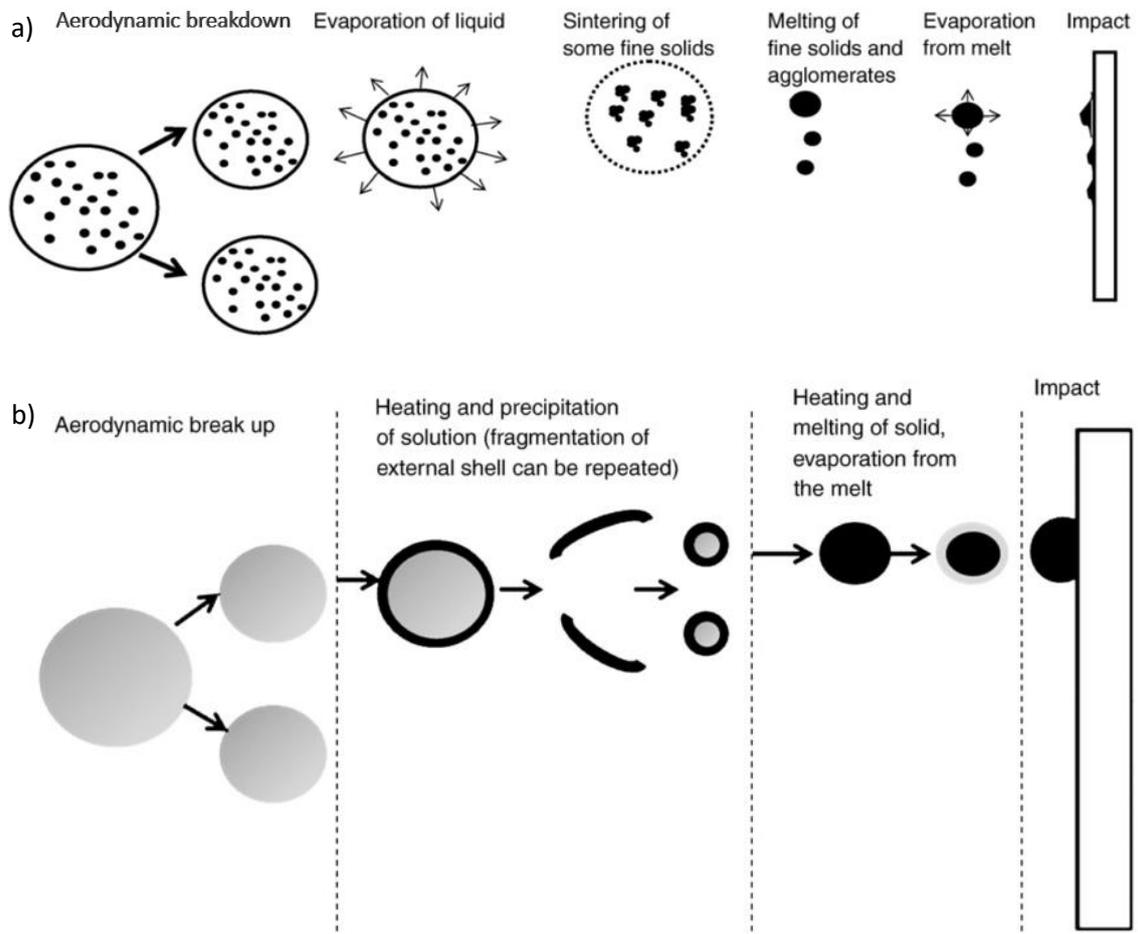

**Figure 3:** *Deposition and particle transformation in-flight for a) suspension thermal spraying and b) solution precursor thermal spraying* (Ref 25).

These techniques increase the flexibility of the plasma deposition technologies, already widely used in the industry, accessing smaller particle size for the feedstock materials, allowing deposited coatings with different microstructures. A field where SPS and SPPS have found great application is the fabrication of thermal barrier coatings (TBC) for high temperature applications. The main reasons are the strain-tolerant columnar structures or vertical cracks and fine porosity on SPS (Ref 31–34) and SPPS (Ref 35–39) deposited coatings, resulting in a thermal conductivity lower that of electron-beam physical vapour deposition (EB-PVD) or traditional APS coatings.

### 2.1.2. Low pressure/vacuum plasma spraying

Plasma spraying in controlled environments was developed in the late 1960s with the aim to reduce the adverse effects arising from the interaction of the in-flight heated particles with the environment. Such detrimental effects include oxidation and undesired contaminations in the coatings. The use of low and very low pressure permits the development of high quality thermal sprayed coatings. The pressures used vary, being commonly in the range of 4000 Pa to 40000 Pa for low pressure plasma spraying (LPPS) and as low as 100 Pa for very low plasma spraying (VLPPS). Any value lower than that is considered as vacuum plasma spraying (VPS). This technique produces coatings with porosity values as low as 1% (Ref 16,40) and columnar structure (Ref 41,42) comparable of those obtained through PVD, of great interest for TBC applications. What is more, it presents the beneficial addition of an increased deposition rate over PVD methods.

## 2.2. Wire arc spraying

Wire arc spraying, also known as twin wire arc spray or electric arc spray, is based on the feeding of two consumable conductive wires (or a core of a non-conductive material on a conductive wire) between which a direct current electric arc is stablished. Once the material is molten, the molten layer is accelerated towards the substrate surface by a stream of atomizing gas. This promotes a further in-flight atomization of the molten particles before their deposition and posterior solidification at the substrate surface (Ref 16,43). The advantages of this deposition method are several. They include the reduced cost of the process, both in terms of equipment and operational costs, being a very cost-efficient deposition technique. It also shows absence of unmelted or semi-molten particles, a high deposition rate when compared to other thermal spray processes and a low thermal transfer to the substrate. All these factors make wire arc spraying one of the less expensive techniques, nevertheless, the particular characteristics of the produced coatings such as high porosity or low bonding strength make its use somewhat limited.

## 2.3. Flame spraying

Flame spray was the first of the thermal spray techniques to be devised, being developed by Schoop around 1909 (Ref 44). The basic principles are still applied to today's modern conventional flame spray guns. The combustion of fuel gases is used to impart heat to the feedstock particles. At the same time, it produces an expanding gas flow which in combination of additional gases creates the required jet applied to accelerate the material towards the surface to be coated. Typical temperatures for this technique are around 3000 K and particle velocities of up to 100 m s$^{-1}$ are usually applied (Ref 16); however, in order to improve the initial design on which flame spraying is based, several variations have been developed with a focus on different flame temperatures and particles velocities.

### 2.3.1. Detonation gun (D-gun) spraying

By confining oxygen and combustion gases such as acetylene within a closed tube and initiating the combustion process with a spark, a high pressure shock wave is created. The shock wave imparts an increased heat transfer and considerable higher kinetic energy to the powder particles, reaching flame temperatures of around 4000 K and velocities of up to 1000 m s$^{-1}$ (Ref 45). The combustion cycle is repeated with a frequency of 3 to 6 Hz to produce a semi-continuous stream of heated and highly accelerated particles. The result is a coating with better adherence strength and reduced porosity to those deposited through the use of conventional flame spray techniques (Ref 16,45,46).

### 2.3.2. High velocity oxy-fuel spraying

High velocity oxy-fuel (HVOF) spraying was developed based on concepts from jet engines, and shares some common features with the detonation gun process. The technique relies on the combination of oxygen and fuel gases inside a combustion chamber, creating a highly pressurised mixture. A small-diameter nozzle is used to direct the gases towards the substrate surface. The combination of elevated pressures, high gas flow and high combustion temperatures produces a supersonic gas jet at the exit, with particles velocities as high as 1000 m s$^{-1}$ and jet temperatures of approximately 3000 K (Ref 47,48). The main differences with detonation gun are the continuous gas stream exiting the nozzle and the free expansion of the compressed flame upon exit at the De Laval nozzle (Ref 49). These factors combined produce coatings with lower porosity and enhanced adherence strength than conventional flame spray coatings (Ref 4,47,50).

#### 2.3.2.1. High velocity air fuel spraying

A variation of the traditional HVOF spraying technology involves the use of air instead of oxygen, giving raise to high velocity air fuel (HVAF) spray. The difference of this method is a reduced flame temperature, due to the lack of a highly exothermic fuel. This has a beneficial effect for feedstock materials with relatively low melting points. It also implies a change in the microstructure and final properties of the deposited coating. In addition, HVAF is less expensive than HVOF, which could represent an advantage for its implementation in the industrial landscape.

### 2.3.2.2. Suspension/solution high velocity oxy-fuel spraying

The use of suspensions and solution precursors as the injection medium has also been developed for HVOF, as well as the already seen plasma spraying technique. The underlying concept is similar, for both suspension HVOF (SHVOF) (Ref 22,51,52), also called high-velocity suspension flame spraying (HVSFS) (Ref 22,53), and solution precursor HVOF (SPHVOF) (Ref 22,51,54–56). They use nano-sized particles to promote the development of coatings with different microstructures in terms of splat morphology and porosity, and therefore different physical properties from conventional flame spraying. Additional research in this field is needed to better understand the potential benefits of SPHVOF and its effect on the microstructure of the deposited coatings.

## 2.4. Kinetic spraying

Kinetic or cold spraying, as the name indicates, is based on the transfer of higher amounts of kinetic energy into the feedstock particles in order to achieve the desired bonding strength upon impingement at the substrate surface. It contrasts with the usual use of heat transfer seen in other thermal spray technologies. This allows for the deposition of deformable, ductile feedstock powder particles without the need for the traditional melting, impact and posterior rapid solidification pathway, effectively reducing the intrinsic residual stresses upon deposition and the in-flight particle oxidation (Ref 50). The basic fundamental of the process is the use of pressurized gases with reduced oxidation potential, such as nitrogen or helium. The gases are moderately heated (up to 1000 K, generally below the melting point of the feedstock particles) in order to increase the gas flow velocities rather than heating the particles themselves (Ref 4). Once the desired pressure and temperature conditions are achieved, the gas is conducted through a De Laval nozzle (Ref 49) which accelerates it to supersonic velocities (up to 1200 m s$^{-1}$) while reducing the gas temperature as it expands. This allows for the temperature to reach in occasions values below room temperature (Ref 3,57). The resulting coatings have the same phase content as the powder feedstock without oxide contamination and low porosity, with preference for compressive residual stresses instead of the usual tensile stress of other thermal spray technologies and low ductility caused by the extensive work hardening involved in the deposition process (Ref 3,50,57).

A different approach is followed in the case of low pressure cold spray (LPCS) which, as its name implies, produces the deposition of the feedstock particles at a lower carrier gas pressure than the common cold spray, or high pressure cold spray. The reduced pressure required in the case of LPCS presents some advantages, such as smaller size and lower cost for the required equipment (Ref 58), making it very attractive for portable, hand-held systems for on-site deposition or repairs. Nevertheless, low pressure equates to lower particle velocity, which mainly affects the deposition efficiency of LPCS, being considerably lower than high pressure cold spray (Ref 59–61).

As a summary, Table 1 and Figure 4 give an overview of the physical conditions for each of the thermal spraying technologies here described.

| Process | Jet temperature (K) | Jet velocity (m/s) | Particle temperature (max) (K) | Particle velocity (m/s) | Coating porosity (%) | Spray rate (g/min) |
|---|---|---|---|---|---|---|
| Atmospheric plasma spray | 15000 | 300 - 1000 | > 4100 | 200 - 800 | 5 - 10 | 50 - 150 |
| | 5800 - 8600 | --- | --- | 240 - 1220 | --- | 80 - 380 |
| | 5300 - 25000 | --- | --- | 240 - 1200 | --- | --- |
| | 5800 - 11300 | --- | --- | 30 - 180 | < 2 | 80 - 380 |
| | 19000 | --- | --- | 100 - 300 | 0.5 - 10 | 8 - 170 |
| Low pressure/Vacuum plasma spray | 12000 | 200 - 600 | > 4100 | 200 - 600 | 1 - 10 | 25 - 150 |
| | 8600 | --- | --- | 240 - 610 | --- | 180 |
| | --- | --- | --- | --- | --- | --- |
| | 11300 | --- | --- | 240 - 610 | < 0.5 | 170 |
| | --- | --- | --- | --- | --- | --- |
| Wire arc spray | > 25000 | 50 - 100 | > 4100 | 50 - 100 | 5 - 20 | 150 - 2000 |
| | 5800 | --- | --- | 240 | --- | 270 |
| | 3300 - 6300 | --- | --- | 50 - 150 | --- | --- |
| | 5800 | --- | --- | 240 | 2 - 8 | 150 - 2000 |
| | 4300 - 6800 | --- | --- | 80 - 150 | 10 - 25 | 17 - 830 |
| Flame spray | 3500 | 50 - 100 | 2800 | 50 - 100 | 10 - 15 | 30 - 50 |
| | 2500 - 3100 | --- | --- | 30 - 180 | --- | 120 - 150 |
| | 3300 | --- | --- | 40 - 100 | --- | --- |
| | 2500 - 3100 | --- | --- | 30 - 180 | 6 - 15 | 120 - 150 |
| | --- | --- | --- | --- | --- | --- |
| Detonation gun spray | 5500 | > 1000 | --- | --- | < 5 | --- |
| | 4200 | --- | --- | 910 | --- | 17 |
| | --- | --- | --- | --- | --- | --- |
| | 3400 | --- | --- | 910 | < 1 | 17 |
| | --- | --- | --- | --- | --- | --- |
| High velocity oxy-fuel spray | 5500 | 500 - 1200 | 3600 | 200 - 1000 | < 5 | 15 - 50 |
| | 3400 | --- | --- | 610 - 1060 | --- | 230 |
| | 3300 | --- | --- | 400 - 800 | --- | --- |
| | 3400 | --- | --- | 610 - 1500 | < 0.5 | 230 |
| | 2900 - 3400 | --- | --- | 550 - 1000 | 0.5 - 5 | 17 - 170 |
| Kinetic spray | --- | --- | --- | --- | --- | --- |
| | --- | --- | --- | --- | --- | --- |
| | 300 | --- | --- | 400 - 800 | --- | --- |
| | --- | --- | --- | --- | --- | --- |
| | --- | --- | --- | --- | --- | --- |

**Table 1:** *Physical conditions and deposited coating characteristics for the different thermal spraying technologies. Data from* (Ref 4,16,43,50,62).

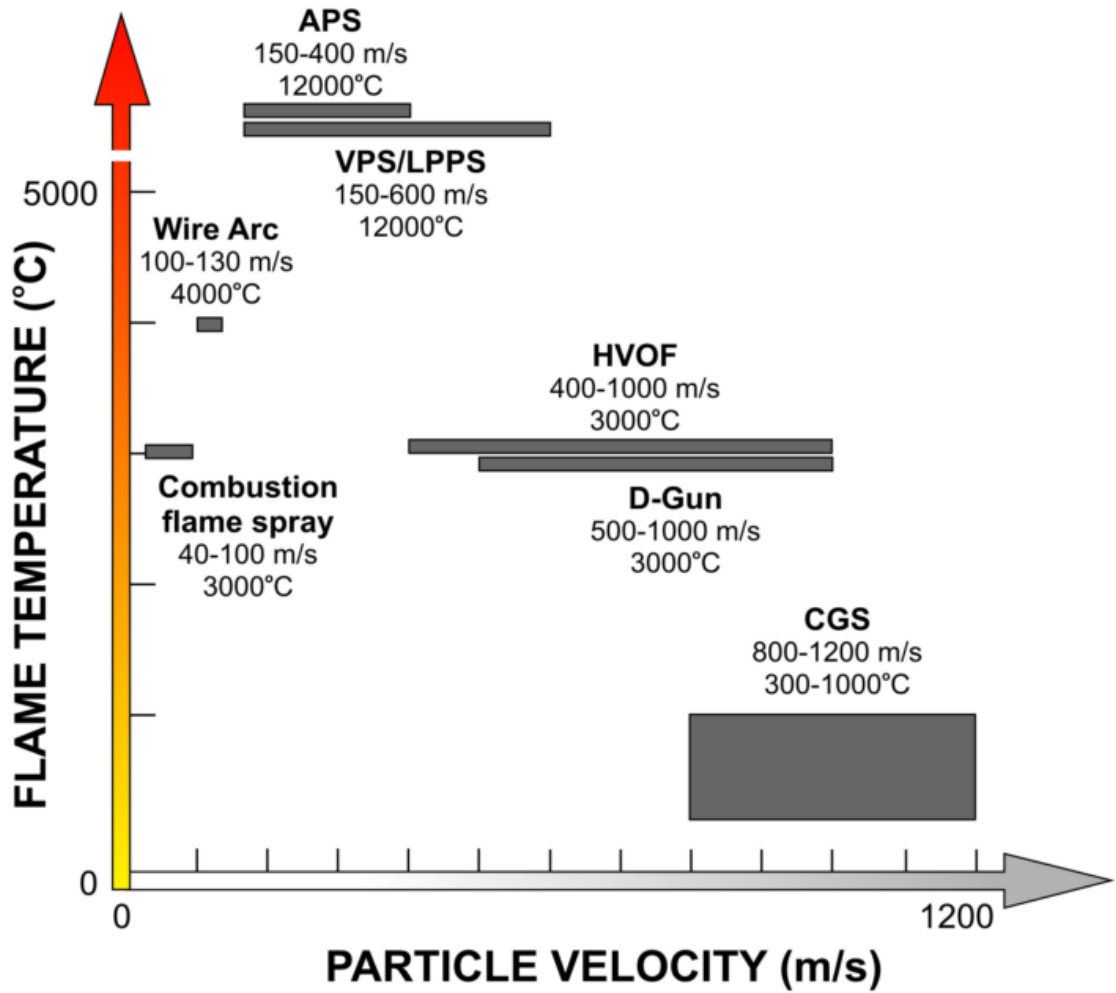

**Figure 4:** *Schematic representation of the typical flame temperature and particle velocity for atmospheric plasma spray (APS), vacuum/low pressure plasma spray (VPS/LPPS), wire arc, conventional flame spray, high velocity oxy-fuel (HVOF), detonation gun (D-gun) and cold gas spray (CGS)* (Ref 48).

3. **Functional Coatings**

The definition of functional coating varies slightly depending on the context on which it is used or the different points of view present in among the experts in the field of thermal spraying; however, in this work a functional coating has been defined as a deposited coating which has a passive, integrated, new functionality beyond the traditional protective capabilities. As such, in this section, the aforementioned definition is applied to classify the functional coatings developed attending to its functionality. For instance, its capability to kill pathogens or prevent infections on orthopaedic implants, the hindering of adhesion and growth of algae and hard shell organisms on water submerged equipment, creation of water- and ice-repellent surfaces or the deposition of coatings with electromagnetic or electrochemical properties.

### 3.1. Anti-microbial

The appearance and adhesion of bacteria and microorganisms onto surfaces cause severe complications such as surgical site infection or chronic wounds in the medical field (Ref 63) or health-related issues due to expired and contaminated products in the food industry (Ref 64). For these reasons the use of anti-microbial coatings, which prevent or hinder the development of noxious microorganisms, has gained popularity in the recent years. To achieve this anti-microbial behaviour several approaches have been taken, being summarised into three main categories. The first one is the creation of an anti-adhesive surface, which prevents the adherence of bacteria and the consecutive formation of biofilms, through physical or chemical modifications. The second approach is the creation of coatings with anti-bacterial agent release capabilities, with highly concentrated and localised doses only where needed. This in turn limits the potential toxicity and resistance development. Thirdly, the use of biocidal (or contact-killing) coatings where compounds with bactericidal activity are immobilised at the surface to provide a continuous protection effect (Ref 65,66). Although all three methods are capable of delivering an anti-microbial effect to the coated surface, each has its own deficiencies that should be considered when designing the application.

In the case of anti-adherent surfaces, the creation of a broad spectrum morphology is complicated due to the non-specificity of the method. A surface with low attachment of a specific bacterial strain might not present the same behaviour with other pathogens, limiting its general application. In addition to non-specificity, anti-adherent surfaces suffer a great functionality loss when wear is present, due to the alteration of the designed morphologies. Despite these deficiencies, the absence of antibiotics or similar agents as the active principle presents a promising approach to prevent the appearance of antibiotic resistant bacteria. As it stands now, this method is mainly applied as a secondary approach in combination with other principles, rather than the main solution against pathogen proliferation (Ref 67).

The loading of an antibiotic or antibacterial agent into the coating has been a popular approach to achieve anti-microbial surfaces, although the method presents one main drawback. The finite nature of an embedded reservoir within the coating implies a time constraint in the duration of the effect. After said time, the reservoir will be depleted and the coating will fail to prevent bacteria proliferation. Despite the severity of this constraint, this method is well suited for applications where a localised and brief delivery of antibiotics is needed. An example would be the protection of implants during surgical procedures and in the following hours, to prevent contamination and infection (Ref 68).

To prevent the shortcomings associated with a limited reservoir, permanent immobilisation of biocidal compounds has been explored. This method, based on the creation of a contact-killing surface rather than relying on a loaded agent, represents a compromise between the lack of specificity and the limited operation time present in the previous methods. The wide array of loaded components or enzymes covalently attached via polymeric chains provide a broad spectrum while the covalent bonds ensure that the components remain fixed in the surface to provide the desired effect (Ref 69). This approach is not free of complications, mainly due to stability issues of the attached components, but it has been the preferred choice for anti-bacterial coatings due to its flexibility and benefits.

Despite the wide use of contact-killing coatings, some considerations should be taken into account when considering thermal spray as the deposition technique. One of the limitations of thermal spraying techniques is that they require heat resistant feedstock materials, limiting the use of most of the chemical species traditionally used to functionalise surfaces, such as poly(ethylene glycol) (PEG) (Ref 70). In view of this limitation, the interest in the development of functional coatings with anti-bacterial properties has been focused on the modification of well stablished thermal sprayed coatings with the addition of a biocidal agent that does not degrade at high temperatures, providing a controlled release over time. With this goal in mind, first a suitable matrix should be available, one that can be efficiently deposited using thermal spray and with a proven record on the medical field. The use of thermal sprayed hydroxyapatite (HA) coatings on orthopaedic implants enjoys a wide acceptance in the medical field since the late 80's (Ref 71), being exhaustively studied since then (Ref 72–77). As reported by Sun *et al.* in the early 2000s (Ref 11), the use of thermal sprayed HA coatings on metal implants presents several advantages and a positive potential for its use in the medical field. This trend has been confirmed by the predominance of literature published between 2010 and 2015 in the biomaterials field on the use of HA coatings on titanium substrates for implant applications (Ref 3). Taking into consideration the status of HA as the standard in thermal sprayed coatings and the decades of research on its behaviour and response to living tissue (Ref 71), the choice of HA as a base material for the development of anti-microbial functional coatings has been, and with good reason, a popular one.

The addition of silver as the anti-microbial agent has been considered on the early studies of functional coatings due its biocidal effect and the difficulty for pathogens to develop resistance to it. The behaviour of HA doped with Ag, with the addition of poly-ether-ether-ketone (PEEK) to form HA-Ag/PEEK coatings, was investigated by Sanpo *et al.* (Ref 78) using cold spray deposition. The lower temperatures involved in this method, as compared to other thermal spraying techniques as seen in section 2, preserves the feedstock powder chemistry and phase composition, as confirmed by EDX analyses. The bactericidal behaviour was reported on *Escherichia coli,* with Ag/PEEK weight ratios varying from 20:80 to 80:20. It was demonstrated the preferential biocidal effect of silver over only PEEK, with biocidal activities 10 times higher in the case of 80:20 HA-Ag/PEEK over the 20:80 HA-Ag/PEEK coating.

The potential of hydroxyapatite combined with silver was also exploited by Noda *et al.* (Ref 79). They reported the use of HA powder mixed with silver oxide, taking advantage of the higher temperatures experienced during flight when using flame spraying (2700 °C) to in-flight melt the powder and form an amorphous calcium phosphate (CP) coating on top of pure titanium substrates. Within the coating there was presence of fully melted, amorphous $Ag_2O$. This approach presents two main characteristics. First, the CP recrystallizes into HA after exposure to simulated body fluid at 37 °C, providing the desired biocompatibility and bone adhesion. Secondly, the presence of silver oxide in the coating, acting as a reservoir and providing a slow release of Ag ions into the environment, provides the required biocidal functionality to the coating. The reported number of viable *Staphylococcus aureus* bacterial colonies after incubation for 24 h was 10000 times lower on Ag-CP coatings with respect to control CP coatings.

Although the predominance of thermal sprayed, doped HA coatings is clear on the field of medical and dental metallic implants, the use of thermal spraying techniques for the deposition of anti-microbial has found other applications. Other base materials such as chitosan (a natural, non-toxic, biodegradable and biocompatible polymer, popular for the development of contact-killing surfaces (Ref 80)) has been doped with recognised biocidal elements such as copper (Ref 81). Cold sprayed coatings of chitosan doped with copper and aluminium (Cu/Al ratio of 75:25 (wt. %)) were tested on *Escherichia coli*. After 24 h a reduction of 22 % on the presence of *Escherichia coli* was reported with respect to control, uncoated glass substrates (Ref 82). The same study was conducted on cold sprayed ZnO nanopowder mixed with Al in weight ratios 20:80, 50:50 and 80:20. They reported enhanced biocidal capabilities for the higher ZnO containing coatings. For such coating the count for *Escherichia coli* after 24 h was 7 times lower than the uncoated glass substrate and coated with just pure Al (Ref 83).

It has already been stablished on the previous publications here presented (Ref 78,82,83) or on the thorough review by Vilardell *et al*. (Ref 84) the benefits of using a deposition technique with lower heat transfer, as in cold spraying, for the deposition of heat sensitive powders. Incidentally, the microstructure produced by cold spraying has been demonstrated to present an increased biocidal activity, three orders of magnitude greater than other higher temperature techniques such as plasma or wire-arc spraying. This effect has been reported by Champagne and Helfritch (Ref 85) on the deposition of pure Cu on aluminium substrates, including a hospital tray entirely coated as a proof-of-concept. The explanation for this noticeable difference on anti-microbial capabilities lies on the work hardening of the copper particles during deposition using cold spraying. This implies a high dislocation density, which enhances the diffusion of $Cu^{2+}$ ions, responsible for the pathogen elimination.

Nevertheless, cold spray is not the only available technique with that characteristic. Wire-arc, also described on section 2, was used by Gutierrez *et al.* (Ref 86) to apply a high content Cu alloy (>60% copper) coating onto medium-density fibre-board (MDF), a popular material in furniture. Their work demonstrated the increased anti-microbial capabilities against *Staphylococcus aureus* and *Escherichia coli* of thermally sprayed Cu coatings, with a lethality ratio 3-4 times higher for the pathogens mentioned in comparison with stainless steel or Cu metal sheets. The microstructure impact of wire-arc sprayed coatings was also investigated by Sharifahmadian *et al.* (Ref 87), reporting the direct correlation between the defects created by the deposition technique, such as grain size, micropores and microcracks and its anti-bacterial properties. The authors suggest that such features promote the release of ions from the surface into the environment, which enhances the antibacterial activity.

### 3.1.1. Biocidal mechanisms of silver, copper and zinc oxide

Several compounds have been introduced in this work as biocidal strategies on functional coatings, and in this section, the mechanism behind the pathogen elimination for each specific case is presented and explained.

As seen before, silver is a recurring choice (Ref 78,79,88–90) for its biocidal capabilities and its effectiveness against strains of antibiotic-resistant bacteria, since no Ag-resistant strain has been found yet (Ref 89). The anti-bacterial mechanism is initiated by the diffusion of $Ag^+$ ions from the surface of the, for instance, HA coating into the surrounding tissue (Ref 88). Once the silver ions are in contact with the bacteria, several mechanisms have been proposed for the biocidal effect, but two main effects seem to be predominant. One, the alteration of microbial DNA, which in turns prevents replication. And two, the disturbance of the bacterial electron transport and respiratory chain, leading to its inactivation (Ref 90,91).

The use of copper for antimicrobial purposes (Ref 81,82,85–87,92–94) dates back to the Ancient Egypt, with descriptions of its use to sterilise chest wounds and purify drinking water (Ref 95). Still, the mechanisms for its biocidal activity are not fully understood yet. As in the case of silver, it has been proposed that the combination of multiple effects is the responsible for the destruction of pathogen cells rather than a single mechanism (Ref 81,92). Nevertheless, the complexity of the problem is increased when different cell lines are considered, as these combined effects also vary for different bacteria (Ref 96). Despite this, different mechanisms have been identified as contributors to the biocidal activity of free copper ions. For instance, the formation of highly reactive hydroxyl radicals (OH$^-$), with damaging effects to bacteria (Ref 94,97), through the change in oxidation state between $Cu^+$ and $Cu^{2+}$ (Ref 81). The substitution of Zn or other metal atoms on binding sites on proteins as also been pointed out, leading to conformation change and the loss of protein function (Ref 92,93) and deactivation of protein by substitution of iron on Fe-S clusters (Ref 98).

An example of the use of ZnO on thermal sprayed functional coatings has been previously presented by Sanpo *et al.* (Ref 83), although its use as an anti-microbial component as nanoparticles and aqueous suspension has been extensively studied. Similar to the use of silver and copper, a complete understanding of the biocidal mechanism of ZnO has not been reached yet, with several effects being proposed in the literature. Three main contributions have been identified, although there is still controversy as to which one represents the main anti-bacterial pathway (Ref 99–104). Firstly, ZnO under illumination with ultraviolet and visible light presents photocatalytic effect (Ref 105–108), described in detail in section 3.1.2, which possesses biocidal capabilities. However, ZnO exhibits a clear anti-bacterial effect even in the absence of illumination. Secondly is the formation of reactive oxygen species (ROS) such as the already mentioned OH$^-$ radicals and $H_2O_2$, which has been reported to inhibit bacteria growth (Ref 109–114). Thirdly, ZnO particles in direct contact with microbial membranes are known to lead to their destabilisation, initiating damage and eventually causing the breakdown of the pathogens (Ref 115–120), although the specifics of this mechanism are not well understood.

### 3.1.2. Photocatalytic effect for anti-microbial applications

Parting from the already presented approach of release-based anti-microbial coatings, the photocatalytic effect provides an alternative method for the development of biocidal surfaces. This method does not rely on embedded agents to provide the biocidal effect, but it also differs from the traditional contact-killing solutions. The photocatalytic effect is based on the illumination of a material, which decomposes compounds by oxidation. As seen in Figure 5, the illumination of the $TiO_2$ coating with photons carrying energy equal or greater than the band gap results in the creation of electron-hole pairs in the titania conductance and valence band, respectively. There is a probability that these charge carriers will transfer or diffuse to the coating surface, where they can interact with adsorbed water and molecular oxygen. The produced electrons usually take part in photoreduction reactions, such as the production of $O_2^-$ radicals, whereas the correspondent holes produce the photooxidation of water molecules, forming free hydroxyl radicals ($OH^-$). These present already mentioned bactericidal effect on numerous bacteria, such as *Escherichia coli* (Ref 121,122) and *Pseudomonas aeruginosa* (Ref 123,124).

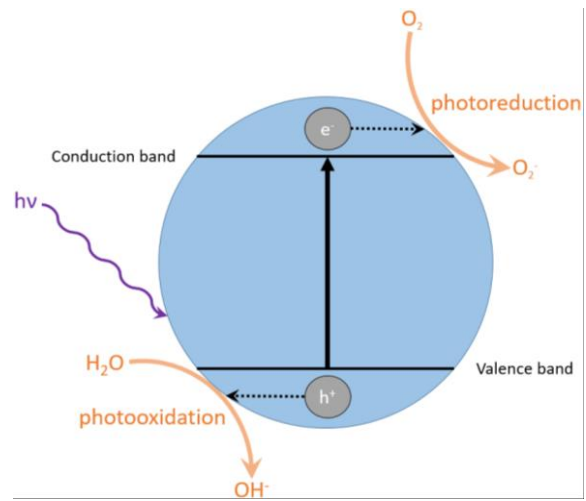

**Figure 5:** *Schematic of the photocatalytic effect in $TiO_2$, with the production of free hydroxyl ($OH^-$) upon illumination. Redrawn from (Ref 125).*

This mechanism has been exploited by George *et al.* (Ref 124) and Jeffery *et al.* (Ref 123), making use of the photocatalytic effect of nanostructured thermal sprayed titania. An initial publication (Ref 123) showed the ability of HVOF-sprayed nanostructured $TiO_2$ to kill up to 24% of *Pseudomonas aeruginosa* deposited after illumination with white light for 120 minutes. Attempting to improve these promising results, George *et al.* (Ref 124) deposited nanostructured and conventional $TiO_2$ coatings using flame spraying, adding Cu as an additional biocidal component, with the aim to take advantage of the potential synergetic effects arising from both the photocatalytic effect and the biocidal nature of copper. The addition of small quantities of Cu (5 wt. %) increased the biocidal capabilities when coupled with white light, due to the production of ROS, which damage the cell wall and membrane, and the photocatalytic reaction under illumination with white light. This effect, although not easily quantifiable due to differences in cells adhesion and cell density between experiments, showed a significant enhancement of the anti-bacterial capabilities attributed to the addition of copper. The authors claimed that the difference in the spraying parameters with respect to HVOF spray, mainly the increased flame temperature and longer in-flight times, produced a disadvantageous phase transformation from the $TiO_2$ powder anatase phase to rutile phase detected in the coating. The anatase phase has been considered the main phase responsible for the photocatalytic effect (Ref 126,127). Nevertheless, recent studies have considered the importance of features such as porosity, the presence of nanostructures or the anatase to rutile ratio, rather than total content of anatase phase. Bai *et al.* (Ref 128) reported the photocatalytic behaviour of SHVOF sprayed $TiO_2$ from rutile feedstock at different flame powers. The coating deposited at lower flame power, with an anatase content of ~20 % and higher porosity levels than the high flame power coatings (being the anatase content ~65 %) surprisingly presented the highest photocatalytic activity, being the current density almost twice as high. The authors suggested the produced bimodal microstructure, with completely melted areas and nano-sized agglomerations, as the cause for the increased activity. The nanostructured formations increase the specific surface area, along with the presence of mixed rutile and anatase areas in which rutile acts as an "antenna", being the electrons captured and then stabilised through transfer into the anatase region, creating high catalytic regions in the interface.

In conclusion, thermal spray methods might not be optimal for the development of contact-killing coatings due to the limitations on temperatures experienced by the feedstock during the process, but they have been proven an excellent solution for others. The possibility to efficiently deposit a robust coating loaded with anti-microbial components represents an excellent opportunity for the medical and dental field. The localised delivery (both in location and time) of biocidal agents is a desired characteristic rather than a drawback for such applications, preventing surgical site infections and extended release of potentially noxious substances. These characteristics, combined with the quick deposition of biocidal components such as copper over large components, i.e. hospital furniture, and the possibilities provided by the photocatalytic effect shown by well-studied components such as $TiO_2$, makes the area of thermal sprayed anti-microbial coatings a thriving one.

### 3.1.3. Membranes for water filtration

Although not a direct anti-microbial application in the sense covered in the previous sections, thermal spray has been applied for the development of membranes for water purification. The ability to deposit a thick film with controlled porosity allows for the design of membranes with tailored mean pore size, effective in the removal of particulates from water. Despite the efforts made by some authors to create ceramic membranes using technologies such as wire-arc (Ref 129), atmospheric plasma (Ref 130) or combustion flame (Ref 131) the manufacturing costs and performance were not comparable to more extended polymeric membranes. An interesting concept was presented in the work by Lin *et al.* (Ref 132) on the development of $TiO_2$ membranes, aiming to combine the filtration capabilities of a ceramic membrane with the biocidal effect of the photocatalytic process in $TiO_2$. Although the concept could lead to new filtration systems that tackle the solid pollutants at the same time as the biological pathogens in water, no experimental measurement on cell strains of the anti-microbial capabilities of the membrane are presented, lacking a strong pillar for the comparison with current filtration systems.

### 3.2. Anti-fouling

The attachment of different aquatic species, classified as soft foulers (primarily algae) and hard foulers (comprehending hard shelled molluscs such as barnacles and mussels) (Ref 133) to surfaces exposed to submersion in water represents a critical factor when considering the efficiency or maintenance cost of any marine and freshwater equipment (Ref 134,135). Of noteworthy consideration is the case of the attachment of fouling to the submerged section of hull in ships, which increases drag forces and the weight of vessels, directly affecting the maximum speed, increasing the fuel consumption and lowering the ship manoeuvrability (Ref 136–139), with a considerable economic impact.

In order to prevent the adhesion and growth of fouling species, anti-fouling coatings are routinely applied. The use of biocidal components on anti-fouling coatings has been the preferred approach; however, ecological considerations limit the use of certain compounds due to their non-specific toxicity, such as tributyl tin (TBT). This compound was worldwide banned for all vessels from 2008 (Ref 139). TBT containing coatings have been replaced by those employing copper (Ref 140), an element already mentioned in this work due to its biocidal capabilities.

Although there are some examples of the use of thermal spray of Cu-based anti-fouling coatings in industrial applications, as demonstrated by a 1984 patent on the deposition of Cu and Cu alloys via thermal spraying (Ref 141), the application of anti-fouling coatings to the hull of vessels has been primarily focused on the use of paints with copper (Ref 142). The reason for the preference of paints resides in the negative interaction between metallic copper present in the deposited coatings and the steel substrate, material widely used for ship hulls. Once copper is deposited directly onto steel, the large potential difference between the two components induces galvanic corrosion (Ref 143).

Despite the fact that the majority of anti-fouling measurements are presented in the form of paints for the reasons previously mentioned, this solution is not optimal for low surface energy thermoplastic polymers (Ref 133,144) such as the polyurethane skins on seismic streamers. Such material presents insufficient adhesion for paints and lack the negative interactions mentioned between steel and copper. For these reasons, the use of thermal spraying techniques, in particular cold spray, has been investigated. Vucko *et al.* (Ref 145) reported an effective technique to provide polymers with an embedded thin layer of Cu, with well-known biocidal capabilities covered in section 3.1.1, using cold spray. Their initial work proved the concept for the deposition of anti-fouling metals into polymers, in this case high-density polyethylene (HDPE) and nylon. The anti-fouling capabilities of these initial results were demonstrated through the complete submersion of the coated samples, and a copper plate as a reference, into the open sea waters of Townsville, Australia. The HDPE Cu-embedded samples prevented biofouling with similar efficiency as the copper plates for up to 250 days, test failed by the nylon Cu-embedded samples. The behaviour of nylon Cu-embedded samples showed reduced anti-fouling capabilities, with 54.3 ± 11.7 % coverage after 181 days in the same conditions as the rest of the samples, while a coverage of 0.6 ± 0.6 % and 0 % was measured for HDPE and copper plates respectively, strongly preventing soft foulers and completely preventing hard foulers. The authors pointed the different depths of embedment of copper particles for the two substrates, being 85.0 ± 1.5 µm for the HDPE Cu-embedded and 40.6 ± 0.4 µm for the nylon Cu-embedded, as the cause. The greater depth of embedment on HDPE allowed for a slow release over longer periods of time, preserving the surface free of foulers for the entire duration of the test. Another study (Ref 146) reported similar results under the same testing conditions, obtained on Cu cold-sprayed polyurethane (PU) samples. This time different spraying parameters (two robot arm lateral speeds) were used in order to elucidate the impact on biocidal capabilities. The faster lateral speed produced a lower density of copper particles, with five time less the amount of Cu per unit area, which were also embedded at a shallower depth, 58.4 ± 1.9 µm compared to 85.6 ± 1.9 µm for the slow speed samples. As a results, the coatings failed to prevent the initial stages of hard foulers after 42 days for the low density PU Cu-embedded samples and after 210 days for the high density PU Cu-embedded samples.

Proved the anti-fouling capabilities of cold-sprayed Cu particles into polymer components, additional research has been carried out to identify the optimal deposition conditions. Lupoi *et al.* (Ref 144) studied the capabilities of unheated, pressurised carrier gas in cold spray, in contrast to the heated conditions previously used, 100 °C (Ref 145) and 400 °C (Ref 146), using a high-speed nozzle design instead to achieve sufficient depth of embedment and surface coverage on different polymer substrates. The authors found out that HDPE substrates suffered from erosion at gas pressures above 2 MPa, achieving optimal depth of embedment and surface coverage using 1.5 MPa, with value of around 35 µm and 57 – 60 % respectively. The addition of computational fluid dynamics simulations to the previous results, allowing to predict the impact velocity of the particles, was reported by Stenson *et al.* (Ref 133). Table 2 summarises the deposition parameters and results achieved on the studies here mentioned.

| Carrier gas temperature | Carrier gas pressure | Scan speed | Substrate | Depth of embedment | Surface coverage | Reference |
|---|---|---|---|---|---|---|
| 100 °C | 2.0 MPa | 100 mm/s | HDPE | 85.0 ± 1.5 µm | 33.6 ± 0.8 % | (Ref 145) |
|  |  |  | Nylon | 40.6 ± 0.4 µm | 60.1 ± 2.2 % |  |
| 400 °C | 2.5 MPa | 33 mm/s* | PU | 58.4 ± 1.9 µm | 21.6 ± 1.4 % | (Ref 146) |
|  |  | 8.3 mm/s* |  | 85.6 ± 1.9 µm | 64.7 ± 1.8 % |  |
| Unheated | 1.5 MPa | 50 mm/s | HDPE | ~35 µm | 57 - 60 % | (Ref 144) |

**Table 2:** *Cold spray parameters for the deposition of copper particles on different polymer substrates, along with the particle depth of embedment and surface coverage achieved. * indicates the lateral movement of the nozzle, and additional rotational speed of the sample was added, with an equivalent lineal speed of 1.77 m/s.*

The use of a high speed nozzle made it possible to deposit the copper particles with unheated carrier gas, preventing potential thermal degradation and deformation of the polymer substrate. However, it also hindered the penetration of particles into the deeper layers, which would have required the use of high carrier gas pressures (2 MPa and above). On the other hand, higher pressures resulted in the erosion of the surface and lower surface coverages. As proved by Vucko *et al.* (Ref 145,146), depth of embedment was the critical factor for longer anti-fouling capabilities and therefore the use of unheated carrier gases limits the effectivity of the coated components. Further studies on the effects of exposure to heated gases for the relevant polymer substrates applied in the marine industry, such as PU on seismic streamer skins, would be beneficial to clarify the potential side-effects and extend the use of thermal spray for anti-fouling applications, being currently limited to niche applications

### 3.3. Hydrophobicity

A hydrophobic surface is defined as having superior water repellent properties, which present several advantages such as reduced contact time with corrosion agents or self-cleaning capabilities, produced by the rolling droplets of water, which carries away dirt. In order to quantitatively evaluate a surface, two parameters are commonly used, namely water contact angle (CA) and slide angle (SA). Contact angle is defined by the tangent to the liquid-vapour interface where it meets the surface, as shown in Figure 6 by the angle θ. The slide angle is the tilt angle required for a static droplet deposited on a surface to start rolling down. As a general definition, any surface with CA > 90° can be considered hydrophobic, while values of CA > 150° and SA < 10° are generally required for a surface to be considered as superhydrophobic.

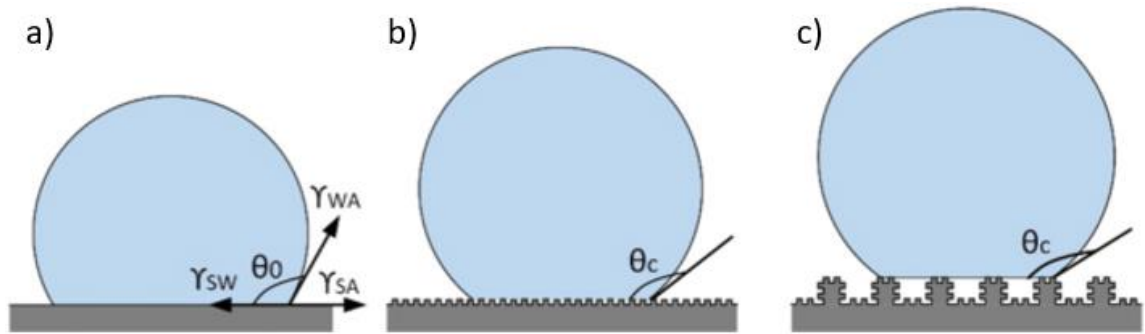

Figure 6: *a) Hydrophobic behaviour of a water droplet on a smooth surface, along with b) rough hydrophobic and c) superhydrophobic surfaces, with their respective contact angles. Schematic b) shows a single-scale/rough morphology, described by the Cassie-Baxter model. Schematic c) shows an improved dual-scale/hierarchical morphology. Adapted from* (Ref 147).

Hydrophobicity has two contributions that explain the particular behaviour presented. First of all, a low surface energy which ensures that the attraction between the water droplets and the surface is minimized. Secondly, a structured surface morphology. As seen on Figure 6, the different levels on the morphology have a direct impact on the hydrophobicity of the surface. Two main models are used to describe different hydrophobic states present on surfaces. The first one, called Wenzel state (Ref 148), assumes the wetting of the entire surface, including in-between the structural features present. On the other hand, the Cassie-Baxter state (Ref 149) assumes that only the upper regions of the rough surface have contact with the liquid, with pockets of air in between such features, as it can be seen in Figure 6(c). Hydrophobicity can be seen in the nature, as in the commonly known case of the lotus leaf (Ref 150) or the wings of certain insects (Ref 151), however, very few are the surfaces that present inherent hydrophobicity features. Therefore, the development of functional coatings that can be easily applied and are readily available for industrial applications have gained substantial interest in the past decades.

The most extended methods to produce hydrophobic materials revolve around the morphological modification of the surface, using techniques such as plasma etching or through the controlled growth of structures using methods such as chemical vapour deposition or lithographic techniques. Another approach is the functionalization of the surface energy using chemical compounds such as polytetrafluoroethylene (PTFE) or poly(ethylene glycol) (PEG) (Ref 152–155). Nevertheless, the use of polymers as hydrophobic coatings has a clear disadvantage on environments where robust mechanical properties are required due to the presence of wear or loads. The application of metals and oxides through the use of thermal spraying techniques allows the production of mechanically sound coatings with excellent hydrophobic capabilities.

Although nowadays the need to address the two contributing factors for hydrophobicity (namely surface energy and surface topography) is generally accepted, the first developments of thermally sprayed hydrophobic coatings focused on topography due to the natural predisposition of thermal spray to produce desirable morphologies. An early example by Teisala *et al.* (Ref 156) presents the development of a nanostructured $TiO_2$ coating deposited on top of paperboard using liquid flame spray with a roll-to-roll set up at ambient pressure, clearly showing the industrial potential of thermally sprayed hydrophobic coatings. Due to the surface morphology, lacking a hierarchical structure, the coating presented an elevated sliding angle. Accounting for the excellent contact angle (up to 160°), the authors proposed the term "highly hydrophobic adhesive surface" instead of superhydrophobic. The reason for this behaviour can be understood considering that the coating is on a Wenzel state, effectively wetting the totality of the coating surface, which increases the adhesive forces, while maintaining a high CA.

Later iterations acknowledged the need for a combined approach between surface energy and topography. A robust and corrosion resistant hydrophobic coating was achieved by Zhang *et al.* (Ref 157) on Fe-based amorphous coating deposited via HVOF. Superhydrophobicity was reached after the addition of a lowering surface energy coating via immersion of the samples on dodecanethiol. An initial study of the as-sprayed Fe-based coating revealed the essential relationship between powder feedstock size and deposition parameters and the final roughness of the coating, which influences its hydrophobic capabilities. The authors stablished that small powder particles and high spraying energy leads to completely molten particles and flatter surfaces (characterised by arithmetic mean surface roughness $R_a$ values of 5.4 ± 0.7 µm), without hydrophobicity characteristics (CA = 71°). The increase in powder particle size and reduction in spraying energy led to higher surfaces roughness, with $R_a$ value between 9.4 and 13.2 µm, and clear hydrophobic capabilities (CA between 120° and 140°). Unfortunately no information on sliding angle of the as-sprayed coatings is given, being only the chemically-treated superhydrophobic surface values reported, with CA as high as 160° and SA around 9°.

Sharifi *et al.* (Ref 147) investigated the impact of the different morphology features present on APS and SPS titania. Due to the use of different deposition techniques, involving aqueous and ethanol suspensions with different dispersing agents, submersion into stearic acid was used to equalize the surface energy of the different samples in order to eliminate the possibility of different surface chemistries, effectively isolating the morphology influence. Their results showed a clear impact of the deposition technique used, with APS presenting SA over 60°. This was attributed to the larger and non-uniform features of APS, with a surface roughness $R_a$ of ~3.7 µm, which prompted a Wenzel state. On the other hand, SPS produced a finer, hierarchical morphology with higher $R_a$ of 6.2 – 6.7 µm described by a Cassie-Baxter state, with CA and SA values of 161° and 1°. Following the success on the reported superhydrophobic surface, a more detailed study of SPS-deposited $TiO_2$ revealed the optimal deposition parameters for improved hydrophobicity (Ref 158). Similar to the conclusion reached by Teisala *et al.* (Ref 156), high contact angles (CA > 150°) were achieved in most of the samples. The sliding angle was the critical factor, with values ranging from over 20°, which can therefore be considered "highly hydrophobic adhesive surfaces" to as low as 1.3° ± 0.3°. This values were achieved on an ethanol-based suspension with 10 wt.% $TiO_2$ deposited on a grit-blasted stainless steel to a surface roughness of 1.5 µm, using 36 kW as the plasma deposition power on with a nozzle diameter equal to 8 mm and a standoff distance of 50 mm. The samples presented excellent superhydrophobic characteristics (CA = 168° ± 1° and SA = 1.3° ± 0.3°) and arithmetical mean height of the surface ($S_a$) value of 8.3 ± 0.1 µm.

As a deeper understanding of the connection between spraying parameters, produced surface topography and hydrophobicity was achieved, attempts to simplify the process were tackled. Still using both morphology and low surface energy modifications, with the difference that only thermal spraying methods were applied, Chen *et al.* (Ref 159) used different deposition techniques to produce hydrophobic coatings. A stainless steel substrate was initially coated with Al using high velocity arc-spray, being further coated with polyurethane (PU)/nano-$Al_2O_3$ using suspension flame spraying. Surprisingly, the arc-sprayed Al coatings showed hydrophilic characteristics with CA below 5°, while the addition of flame-sprayed nano-$Al_2O_3$ alone (without the presence of PU) further maintained the hydrophilic behaviour, with CA below 5°. The marked hydrophilic nature of nano-$Al_2O_3$ is considered responsible for this effect, despite the presence of hierarchical morphology at the surface. The combined effect of a rough surface morphology (although no surface roughness values are reported) with a lowered surface energy induced by the addition of 2 wt.% PU, allowed for superhydrophobic CA values over 150° and SA values of 6.5°. An additional benefit of the superhydrophobic coating was an excellent corrosion resistance, demonstrated by its good electrochemical behaviour on 3.5 wt.% NaCl aqueous solution at room temperature. Following on the synergetic effects of morphology and surface energy, a refined system was achieved by the introduction of a dual-scale, or hierarchical, morphology (Ref 160). The use of a mesh as a micro-patterning plate introduced cone shaped microstructures with finer nanoroughness (no $R_a$ values are reported) on $TiO_2$ APS-coated samples. Once more, the surface energy was lowered depositing a thin film of PTFE/nano-Cu using suspension flame. The reported values for the CA and SA of the surface were 153° and 2° respectively, with the benefit of being a mechanically robust and easy to repair coating.

Despite the fact that the developments allowed to tailor the surface morphology and energy treatments within the realm of thermal spraying, they ultimately relied on the chemical modification of the surface to achieve superhydrophobic behaviours. A recent development made by Cai *et al.* (Ref 161) further increases the possibilities for the superhydrophobicity behaviour of functional coatings, without requiring an additional chemical modification of the surface energy. Their work used rare earths, specifically ytterbium nitrate pentahydrate (Yb(NO$_3$)$_3$)-5H$_2$O, with inherent hydrophobic behaviour accounting for their electronic structure (Ref 162,163). The deposition was done using two mediums, only distilled water and 50% water 50% ethanol, using the SPPS technique. This allowed to directly produce Yb$_2$O$_3$ superhydrophobic coatings. The combined hydrophobic capabilities, due to the chemical properties of the coating, and the hierarchical structures obtained through the use of SPPS produced superhydrophobic surfaces with contact angle 165° ± 2° and roughness values of $R_a$ = 0.9 μm. The elimination of the chemical functionalization of the surface, and the high deposition rates of the thermal spray technique, represents a very interesting candidate for industrial implementation on large structures. Further investigations have been carried out by Xu *et al.* (Ref 164) using the same deposition technique and materials, producing superhydrophobic coatings with SA values of ~163° and sliding angle ~6.5°. Following this rationale, Bai *et al.* (Ref 165) present the use of another cost-efficient deposition technique, SHVOF spraying, to deposit a different rare earth oxide (CeO$_2$). They produced robust, near-superhydrophobic coatings with CA values ranging between 134° (for Al alloy substrate) and 146° (for stainless steel substrate) with surface rough height $S_A$ of 3.6 μm, although no sliding angle results are presented.

### 3.3.1. Icephobicity

A natural extension of surfaces with the capability to repel water droplets is the consideration of similar mechanisms that prevent the formation or adherence of ice. The development of icephobic surfaces represent a field of interest with multitude of applications such as aerospace structures, solar panels or wind turbines, where ice causes a loss in efficiency and an increase in costs. Nevertheless, the mechanisms involved in the prevention of ice accumulation are more complex than those encountered with water in liquid form, creating a thriving field aiming to determine the fundamental interactions (Ref 166–168). On a first consideration, ice needs to be prevented to form in the surface, and if formed, the adhesion strength to the coating should be less than the uncoated equivalent, facilitating its removal. In addition, the different routes involved in the formation of ice and the several solid configurations that water can experience, such as frost, glaze, rime, snow or ice (Ref 166,169), add an additional complexity layer to the problem. Sojoudi *et al.* (Ref 166) present the main differences between the mechanism behind superhydrophobic surface and those involved in the development of icephobic (or pagophobic as they prefer to called them, from the Greek word "pagos" for ice). Figure 7 gives an overview of the main mechanisms considered when designing an icephobic surface. Here, three parameters play the key role. First, topography, which determines the presence of a Wenzel or Cassie-Baxter state, as explained in the previous section. Secondly, elasticity, being a common example the use of silicon for ice cube trays accounting for its flexibility and low surface energy. Thirdly, the liquid extent, where a micro/nanoporous material is infused with lubricant liquid with a low freezing point, providing a smooth liquid interface that reduces droplet retention and ice adhesion strength.

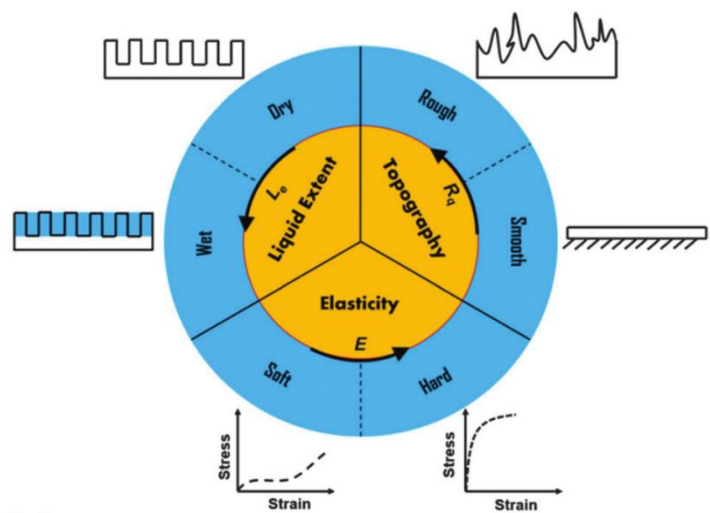

**Figure 7:** *Classification of the three main mechanisms applied in the development of icephobic coatings* (Ref 166).

The use of thermal spray technology to produce icephobic coatings is limited for several reasons. The surface roughness exploited in most thermal sprayed hydrophobic coatings can be detrimental under certain conditions such as high humidity (Ref 170,171). This makes the use of chemical functionalisation of smooth surfaces and liquid extent pathways the preferred alternatives. Noteworthy is the work of Koivuluoto *et al.* (Ref 172) on the deposition of polymers (specifically polyethylene) using flame spray. Their results emphasise the difference with hydrophobic coatings, with a reduction on the ice adhesion strength on the as-sprayed coatings and after polishing of 20 – 25 %. When compared with uncoated, polished substrates, the thermal sprayed coatings presented an ice adhesion strength ~5 times lower than stainless steel and ~7 times lower for aluminium substrate.

The application of new mechanisms already proven in hydrophobic coatings, such as the use of rare-earth compounds with intrinsic hydrophobic behaviour even on smooth surfaces, presents a new field to be investigated which could potentially increase the presence of thermal sprayed anti-icing coatings.

### 3.4. Electromagnetic and electrochemical properties
#### 3.4.1. Electromagnetic properties

Thermal sprayed electrical and magnetic functional coatings comprises the use of materials deposited using thermal spraying techniques with a focus on its electromagnetic properties instead of its chemical properties or as passive barriers. This gives rise to a wide range of applications such as integrated circuits. The importance of these materials is easily understood if the role of thick-film-based electronics in today's world is considered, being the base of the micro-electronics industry. Such importance is demonstrated by a patent by Prinz *et al*. (Ref 173) registered on 1994 on the development of electronic packages by thermal spray. They created electronics structures through the application of masks, building up the different components on each spray run. The use of thermal spraying deposition, capable of efficiently and cost-effectively deposit continuous layers or patterned designs of conductive and isolating materials has gained interest over the last decades. In particular, the desire to produce patterns with linewidths as reduced as possible has driven the investigation of novel deposition techniques within the thermal spray field. Nowadays two different trends can be identified in the development of thermal sprayed micro-electronics.

The first one, called additive-only, relies on the miniaturisation of the thermal spray equipment itself and the use of dynamic apertures at the nozzle (Ref 174). This achieves a smaller flame capable of directly depositing features with a range of thicknesses between 250 μm and a few millimetres. This technique has been successfully applied in the development of electromagnetic shields (Ref 175), antennas for unmanned aerial vehicles (Ref 176) or gas sensors on heat-sensitive substrates (Ref 177). The approach has several advantages present on traditional thermal spray techniques, such as the capability to deposit on heat-sensitive substrates, the flexibility of choice in deposited materials and the readily available infrastructure in the industry. An extensive review on its use and applications accounting to the different properties exploited, such as dielectricity, conductance, resistance, magnetism or superconductivity was written by Sampath (Ref 12). In his work, an analysis of the microstructure and intrinsic physical properties of thermal sprayed functional materials is first presented, with focus on how to understand and control said properties. The different devices that have emerged from the application of electromagnetic thermal sprayed materials are then outlined, with examples of multi-layered circuits, antennas, ohmic contacts and sensors, to name a few. The review concludes identifying the key factors for the limited spread of thermal spray for electric and magnetic applications, mainly the poor understanding of the material properties and the lack of precise enough tools to achieve the required miniaturisation.

The second approach involves the use of machining methods, such as laser micromachining (Ref 178), on a thermal sprayed layer to form the desired pattern as shown on Figure 8. It is therefore labelled additive-subtractive, for obvious reasons. The addition of a second processing step, although it pushes the linewidth down to 15 – 20 µm, considerably increases the time required to fabricate the device. Another effect is the added cost due to the need of a parallel machining infrastructure, when compared to a single thermal spraying unit as in the case of additive-only. Nevertheless, its capabilities have been exploited in the development of thermopiles for power generation or temperature sensing (Ref 179,180), embedded microheaters (Ref 181) and strain gauges (Ref 182). As proven by the recent work in the field, the additive-subtractive approach represents a great technique with potential in the field of prototyping and specialised components, where longer manufacturing times and increased costs per unit are not as critical as in large-scale production.

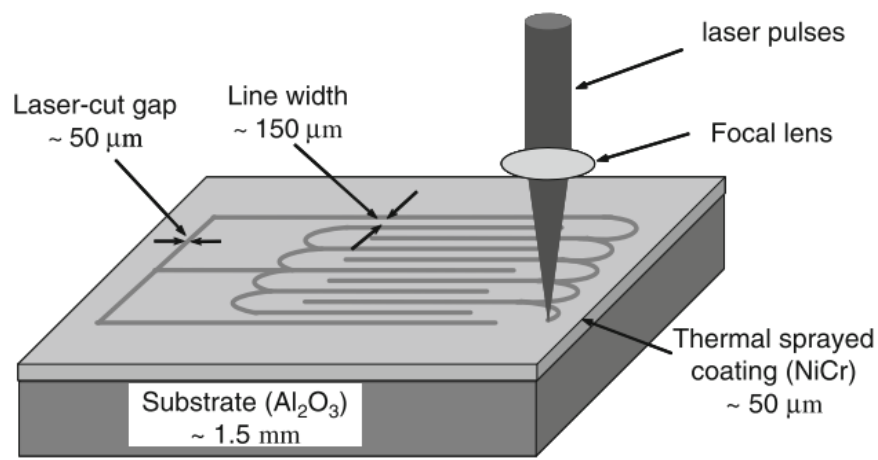

**Figure 8:** *Schematic of the additive-subtractive method used to produce precise patterns through the addition of laser micromachining to a thermal sprayed layer* (Ref 12).

Despite the advancements made towards miniaturisation of thermal sprayed components, their uses are still limited to laboratory scale and cannot compete with the current, well-stablished methods used in the electronics industry. Nevertheless, the production of electrical circuits is not the only application where thermal sprayed materials can be of use. Plenty of modern devices require components of reduced size and suitable electrical properties. For instance, the ability of thermal spray to produce dense, thick ceramic films is more desirable in cases such as piezoelectric components for micro-devices, presenting a clear advantage over other competing methods such as chemical vapour deposition.

In addition to the capability to deposit thick ceramic coatings, another crucial advantage when considering the use of thermal spraying for electrical applications is the confinement of the high temperature process to the actual deposition. This translates into a very low heat transfer to the substrate. On the other hand, when developing multi-layered structures through alternative techniques, such as traditional ceramic coating production (e.g. tape casting, doctor blade) the temperatures required to fire the ceramics might prove too high for the rest of the components, or it could induce undesirable chemical reactions. This implies an investment on high temperature equipment, high melting point substrates and protective layers to avoid the reactions are needed, further increasing the manufacturing costs when compared with thermal spray deposition.

Despite the lower heat transfer present in thermal spray, another factor to be taken into account is that melting of the feedstock particles may lead to the undesirable phase transformation, demanding post-deposition heat treatments if the initial microstructure is to be recovered. Therefore, research in this field has been predominately focused on kinetic deposition techniques, such as cold spray, described on section 2.4 or aerosol deposition (AD). The underlying working principle of such techniques is the ejection of submicrometre powder particles with the aid of a carrier gas. The aerosol flow is directed towards the substrate, being kept under vacuum during the whole process. Since the bonding mechanism relies on the adhesion of the fine particles, these techniques do not alter the feedstock phase composition or suffer from heat induced damage to the substrate (Ref 183). The goal in these cases is to produce a dense layer with unaffected electrical properties.

Considerable success has been achieved in the use of those techniques to develop electrical devices, such as piezoelectrics (Ref 184,185), microwave filters for radio-frequency applications (Ref 186), electro-optic complex films (Ref 187), electrodes for solid state lithium batteries (Ref 188) or $TiO_2$ films for dye-sensitized solar cells (Ref 189,190). In all the examples presented above, kinetic spraying allowed for a higher deposition rate than comparable thick-film deposition techniques such as screen printing (Ref 191), with the benefit of an absence of high temperature exposure, characteristic of other thermal spray techniques. A comprehensive overview by Akedo (Ref 192) covers the applications of kinetic spraying, or room temperature impact consolidation (RTIC), for microdevices. His work presents a thorough comparison of the challenges and opportunities in the fabrication of micro-actuators, requiring dense and thick films of piezoelectric or electrostrictive materials, using conventional methods such as sol-gel or chemical vapour deposition and RTIC. The higher deposition efficiency of thermal spray methods and the advantages mentioned related to the higher control over the microstructure using RTIC, proved the potential of this approach.

### 3.4.2. Electrochemical properties

The advancement in electrochemical cells, both for the conversion of energy and the electrolysis of water, has been an active research topic for the past decades. The specific environment to which the components are exposed (high temperatures, corrosive chemicals and varying electrical current) makes the electrodes a particular interesting point for the improvement and development of better performing electrochemical cells. In addition to an outstanding physical and chemical stability, the electrodes need to have surface area as high as possible, aiming to increase the accessibility of the electrolyte to the electrode surface and to allow an enhanced detachment of the gases produced in the process, which effectively increases the electrocatalytic activity (Ref 193).

When used on power-consumption mode, for instance, to perform alkaline water electrolysis, electrochemical cells are regarded as one of the prime future techniques for the mass production of hydrogen. The process is not new and has been extensively investigated. This includes the use of thermal spray techniques for the fabrication of electrodes, as proven by a 1977 patent (Ref 194). Nevertheless, the ever-increasing demand for hydrogen, mainly due to its applications as clean fuel for transportation powering systems, has boosted the interest for more cost-effective systems. The use of thermal spray techniques has been proven as an effective way to produce electrodes, achieving an excellent control of the microstructure and surface properties due to the flexibility on deposition parameters and different techniques, as shown in Table 3.

| Process | Year | Material | Cell temperature | Reference |
|---|---|---|---|---|
| Atmospheric plasma spray | 1983 | Ni | 80 °C | (Ref 195) |
| Vacuum plasma spray | 1995 | Ni-Al | 70 °C | (Ref 196) |

| Low pressure plasma spray | 1995 | Ni-Al-Mo | 70 °C | (Ref 197) |
|---|---|---|---|---|
| Wire arc spray | 1999 | Ni-Al | 25 °C | (Ref 198) |
| Vacuum plasma spray | 2004 | Ni-Al & Ni-Al-Mo | 25 °C | (Ref 199) |
| Wire arc spray | 2007 | Ni & Ni-Ti | 25 °C | (Ref 200) |
| Atmospheric plasma spray | 2013 | Ni-Al | 30 °C | (Ref 201) |
| Atmospheric and Suspension plasma spray | 2016 | Ni | 25 °C | (Ref 202) |
| High velocity oxy-fuel spray | 2016 | Ni | 25 °C | (Ref 203) |
| Cold spray | 2016 | Ni | 22 °C | (Ref 204) |

**Table 3:** *Compilation of application of thermal spray techniques for the deposition of electrodes for alkaline water electrolysis. Adapted from* (Ref 193).

A higher cell temperature benefits conductivity and the detachment of hydrogen bubbles produced, increasing the electroactivity of the electrode. It carries, however, unwanted effects, such as accelerated corrosion and higher consumption. For these reasons, a carefully designed surface topography presenting multi-hierarchical structures has been favoured to increase the surface area allowing for a reduction in the operating temperature.

Alternatively, electrochemical cells can be used for the generation of electricity, as in the example of solid oxide fuel cells (SOFCs). Different challenges are present for the development of SOFC components, as the configuration is not the same. Firstly, the elevated temperatures needed for the efficient operation of SOFCs (generally above 600 °C) requires a different set of materials than those used for alkaline water electrolysis. Secondly, the design of SOFCs requires a thin, gas-tight electrolyte that provides effective insulation between the two electrodes, while permitting the conduction of oxygen ions, along with micro-porous electrodes that allow for the fuel and air to penetrate. For these reasons, ceramics have been the material of choice, including perovskites such as lanthanum strontium manganese oxide for the cathode, YSZ for the electrolyte and a mixture of Ni and YSZ for the anode (Ref 205,206).

Traditional fabrication methods for those ceramic components present several disadvantages that thermal spray could help to overcome. The high temperature required for the firing process, for instance, presents a bottleneck on the mass production of SOFCs, greatly increasing their cost. It also implies problems such as thermal mismatch stresses on large components, side reactions between adjacent layers and the difficulty of using low-melting materials in the multi-layered structure (Ref 207). The increase time efficiency of thermal spray techniques such as plasma spray and the reduced heat transfer to the substrate (allowing the use of cheaper materials) arise as potential solutions to the previous problems, meaning that thermal spray could drastically reduce the cost of current SOFCs.

Regarding the deposition of electrodes, plasma spray, a common deposition technique in the industry, usually presents porosity levels between 5 – 15 vol.%, which is still far from the 40 vol.% required for optimal gas diffusivity in electrodes (Ref 207). The use of porous metal substrate (Ref 208), the inclusion of larger particles (Ref 205) or with higher melting point (Ref 209) and the addition of "pore formers" (material burned out after deposition, leaving empty spaces in the coating) (Ref 210) have been successful techniques applied to increase porosity.

In the case of the electrolyte, porosity would have a detrimental effect on the cell efficiency and hence a pore free structure is required. Thicker coatings could be applied to minimise the effect of open porosity, increasing the diffusion path; however, this approach has a negative impact on the efficiency, and alternative routes to minimise porosity and defect density have been pursued. For instance, a thorough understanding of the plasma spraying deposition parameters has been followed, in order to determine the optimal conditions for the deposition of crack-free coatings. This allows qualitative low levels of porosity (no porosity measurements are reported) on relatively thin coatings (less than 40 µm) with performance comparable of sintered components (Ref 211).

The production of cheaper SOFCs for electricity generation would represent a major breakthrough for the industry, leading to a greater industrial uptake. Nevertheless, to achieve this, the above-mentioned design requirements should be met, such as porous electrodes or thin, defect-free electrolytes.

### 3.5. Resistive heating

#### 3.5.1. Development of thermal-sprayed coating systems for heating purposes

This section provides a summary of the reported research findings in the area of development of functional thermal-sprayed coatings for heating applications. Different materials and spraying processes that have been investigated by various researchers for fabrication of heating systems and the main contributions of each study in advancement and evolution of the thermal-sprayed heaters are reviewed in this section.

The usage of thermal-sprayed coatings as electrical resistance heating systems has received increasing attention in recent years. However, the application of thermal-sprayed coatings in electronics was proposed around five decades ago when planar ferrite microwave integrated circuits (MICs) were successfully fabricated by using arc-plasma spraying (APS) process (Ref 212). Smyth and Anderson (Ref 213) studied the effect of spraying parameters on the physical and electrical properties of the produced film resistors. It was shown that APS process can be used as an alternative method for producing electrical components and circuit. It was found that a film with sheet resistivity from 5-500 $\Omega$/sq could be obtained by using a mixture of NiO and $Fe_3O_4$ powders with a particle size range of 1 – 20 µm and that APS can be used for cheap quantity production of resistors and conductors with satisfactory long-term stability (Ref 213).

Sampath *et al*. (Ref 214), investigated the application of thermal spray techniques in fabrication of meso-electronics. It was found that multilayer deposits of ceramics and metals that have appropriate electrical properties can be produced by thermal spray methods to fabricate electrical components ranging from insulators, conductors, and resistors. Fabrication of resistor was conducted by deposition of NiCr alloy over alumina by plasma or HVOF processes. The sheet resistance of the fabricated resistor was reported to be in the range of 17,000-54,000 $\Omega$/sq. Low cost, high production rate, flexibility of use of thermal spray process, and capability of producing millimetre-thick layers of conductors and insulators made this process an alternative for fabrication of components for power electronics (Ref 214). Similarly, deposition of ceramic feedstock particles by using HVOF process that is accompanied by oxygen-rich flames and the use of cold spray process have been found to be appropriate methods in synthesis of high-quality dielectrics and conductors, respectively (Ref 50).

In an early effort (Ref 215), air plasma spraying process was used to fabricate the resistance heaters. The metal film heater (molybdenum), which was isolated from the faceplate by a plasma-sprayed ceramic film, was able to generate heat fluxes up to 7.2 MW/m$^2$ over an area of 10.3 cm$^2$ (Ref 215). Therefore, it was shown that thermal-sprayed coatings can be used as heating systems. Given the direct deposition of the films atop the test surface, as opposed to mechanically attaching the heater to the surface, the thermal contact resistance between the thermal-sprayed heating element and the substrate is minimised (Ref 216). This results in fabrication of more efficient heating elements. The structure of the heating system depends on the electrical resistance of the metallic alloy that is required to be achieved, the substrate material, and its electrical conductivity. Owing to the high thickness required to achieve the appropriate electrical resistance, thermal spraying techniques are preferred over vapour deposition for fabrication of the coating-based heating elements (Ref 216). It is well-established that for the cases in which the substrate is electrically conductive, an intermediary ceramic layer, which possesses dielectric properties, is required to prevent short circuiting and leakage current in the heating system. Therefore, it is required to develop a multi-layered coating system that consists of the two main elements, namely the heating element that is usually a metallic alloy and an electrically insulating ceramic layer.

### 3.5.1.1. Deposition of the ceramic coating as an electrically insulating layer

Alumina is widely used as an electrical insulator due to the dielectric properties that this material possesses. Given the rapid and simple coating of large surfaces by using plasma spraying process, deposition of electrically insulating coatings have been an appealing option in electronics industry. It has been shown that the volume resistivity of plasma-sprayed alumina coating can be very high within the range of $10^9$ - $10^{10}$ Ω cm (Ref 217). It has been reported in the literature that the electrical resistivity of alumina depends on several parameters including applied pressure, humidity, microstructural characteristics, and phase composition of alumina (Ref 218–222).

Luo *et al.* (Ref 218) demonstrated that the surface resistivity of alumina coating decreases by applying pressure on it, but it still was larger than $10^6$ Ω cm when it was compressed at pressures up to 250 MPa. High humidity levels also affect resistivity of the alumina coating (Ref 219–221). In a study by Toma *et al.* (Ref 219) a dramatic decrease about five orders of magnitude in the resistivity of alumina coating was observed when the relative humidity level was increased to 95%. The deterioration of the insulating properties of alumina coating with increasing humidity was explained by increase in the surface conductance of the oxide layer because of absorption and accumulation of water molecules on the coating surface (Ref 220).

Furthermore, the different microstructural features of the coatings due to use of different thermal spray processes affect the electrical resistivity of alumina layer. It was found that the electrical resistivity of the alumina coating that was obtained from HVOF process was slightly higher than the one that was fabricated by APS process, which was likely due to the lower open porosity and presence of defects (Ref 219). Furthermore, it was shown that the alumina that is sprayed by suspension HVOF (SHVOF) has better electrical resistance stability than the alumina coating deposited by the conventional HVOF process due to the retention of higher content of α-alumina (Ref 221).

The phase transformation of alumina is also another well-known, but neglected factor that can negatively affect the electrical properties of the thermally sprayed alumina coatings. The deterioration of electrical properties of alumina is due to the phase change from α-alumina (corundum) to γ-alumina, which is a metastable phase with undesired properties (Ref 221,222). The phase change is attributed to the rapid cooling of the molten alumina particles below the temperature at which the atomic rearrangement can occur before the stable phase of corundum is reached (Ref 223). It was found by McPherson (Ref 224) that the metastable form is retained during cooling of particles that are less than 10 μm diameter and the particles larger than that may transform to α-$Al_2O_3$ during the solidification.

It has been shown that plasma-sprayed deposits of cordierite (2MgO-2$Al_2O_3$-5$SiO_2$) can also be used as the electrically insulating layers due to the high electrical resistivity and chemical durability of this material (Ref 225,226). Alumina-titania coatings are other alternatives that can be used as electrically insulating layers. In addition to the dielectric properties, they benefit from enhanced toughness, which increase their performance and durability (Ref 12); however, the electrical resistivity of the alumina-titania coatings is partially compromised. As an example, it was observed by Golonka and Pawlowski (Ref 227) that the surface resistivity of arc plasma sprayed alumina was $10^9$ Ω cm, but that of alumina-2wt. % titania was $10^7$ Ω cm.

The main challenge associated with the coating-based insulators is the dielectric breakdown mechanism of the coating, which is indicative of the failure of the electrically insulating layer, and is caused by electrical discharge when high local fields are produced. This leads to physical degradation of the ceramic layer by formation of a complete failure path. It has been found that voltage at which the breakdown of the APS-sprayed $Al_2O_3$-13% $TiO_2$ coating, which is one of the compositions of alumina-titania systems that is widely used as an electrical insulator, occurs is linearly related to the thickness of the coating. In this respect, the average dielectric strength of the coating was measured to be 20 kV/mm (Ref 228).

It was observed by Killinger *et al.* (Ref 229) that the two most important factors regarding the electrically insulating layer are breakthrough voltage and leakage current, which depended heavily on the internal morphology of the coating that was related to the thermal spraying process that was used for deposition of the coating. It was concluded that the porosity of the deposited alumina coating had a profound impact on the breakdown voltage of this electrically insulating material. As a result, the maximum applicable voltage during operation was noticeably higher for the alumina coating that was deposited by HVOF method compared to the one deposited by APS process (Ref 229).

### 3.5.1.2. Deposition of the metallic alloy coating as the heating element

A material that is electrically conductive and has high electrical resistivity is required to be deposited onto the intermediary ceramic layer with dielectric properties to fabricate the heating element. Deposition of several different thermal-sprayed metal alloys, namely molybdenum (Mo), nickel (Ni), nickel-20 wt.% chromium (Ni-20Cr), nickel-5 wt.% aluminium (Ni-5Al), iron-13 wt.% chromium (Fe-13Cr), and iron-chromium-aluminium (FeCrAl), by using various thermal spraying processes including APS, vacuum plasma spray (VPS), high-velocity oxygen fuel (HVOF), wire arc, combustion spray, and wire flame spray, has been reported in the literature for usage as the heating elements (Ref 181,215,216,229–236).

Michels, *et al.* (Ref 216), used three thermal spray techniques, namely air plasma spray (APS), VPS, and HVOF, to produce resistance heating elements that can generate uniform and easily controlled flux. It was found that HVOF and VPS processes compared to the APS could manufacture consistent coatings with physical properties that are close to the bulk materials. In this study, a metallic layer (nickel-chromium alloy Ni80-Cr20), which served as the heating element, was deposited onto a ceramic layer (alumina) that served as the electrically insulating layer. The layers had a thickness in the range of 75 to 300 µm. The configuration of the studied resistive heating system is shown in Figure 9.

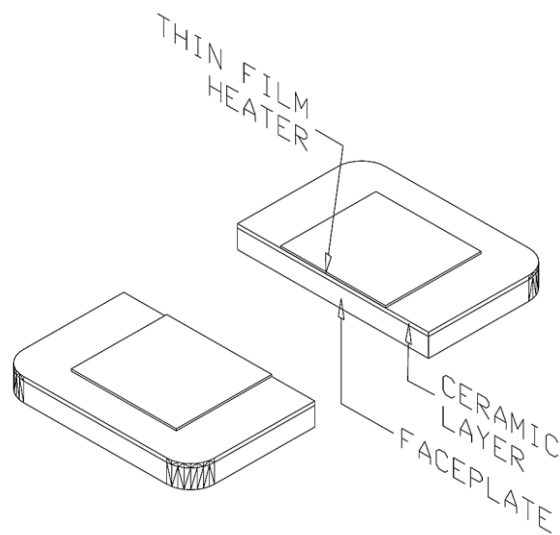

**Figure 9:** *Configuration of heater and insulator films* (Ref 216).

Michels *et al.* (Ref 216) found that the thickness of alumina that was required to isolate the heating element from the substrate was around 100 μm for HVOF, but 200 μm for APS. Furthermore, the highest fluxes that could be generated from the fabricated heating system before failure was as high as 10.6 MW/m$^2$ and 17.2 MW/m$^2$ for HVOF and VPS films, respectively. The heaters failed at very high level electrical currents. It was found that the failure mode of the thermal-sprayed heating system was delamination of the Ni-Cr layer from the insulating ceramic layer. Therefore, the evolved thermal stresses between the deposited ceramic and metallic films due to the mismatch between the thermal expansion coefficients were suggested as the root cause of the failures (Ref 216).

Li (Ref 231) studied the electrical properties of both essential components of the heating systems, namely the insulator and the film heater. The nickel-chromium (Ni20Cr) coating was deposited onto the alumina coating by plasma spraying process and using spraying mask technique to form the metal-ceramic multilayer composite that is required for fabrication of the heating devices. The layout of the developed heater and the micrograph taken from the cross-section of the coated sample is shown in Figure 10. The thickness of the alumina and NiCr coatings (see Figure 10(b)) were measured to be 200 and 30-50 μm, respectively. It was found that the tensile residual stresses reach very high values at the surface of the NiCr coating and deposition of thicker coatings (more than 50 μm) would lead to delamination of the metal film.

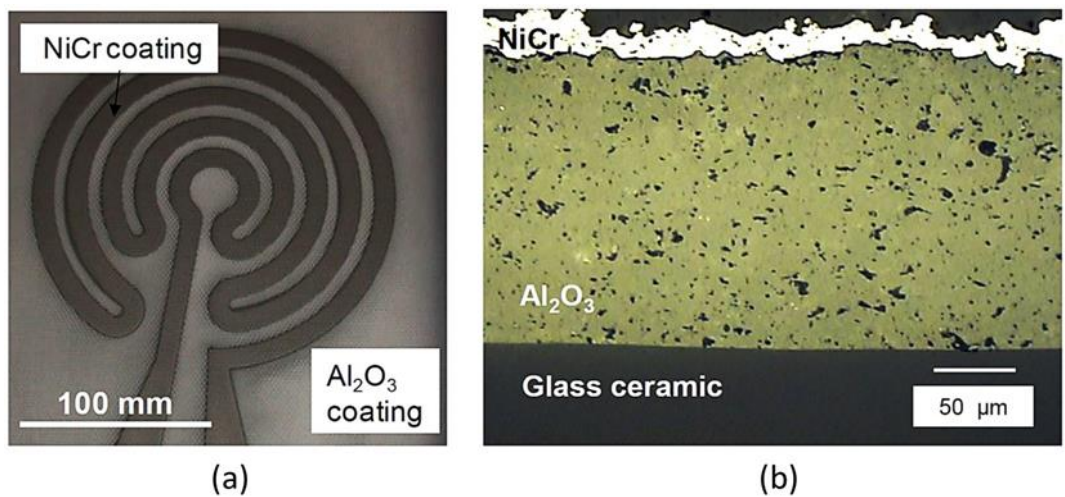

**Figure 10:** *a) Layout of the fabricated heater and b) the micrograph taken from the cross section of the coating system* (Ref 231).

Floristan *et al.* (Ref 232) investigated the development of plasma-sprayed electrically conductive coatings on glass ceramic substrates. Given the low or even negative coefficient of thermal expansion (CTE) of glass ceramics, they are suitable materials for production of cook top panels.

This enables this material to be in contact with cold objects (e.g. pots from the fridge) while they are operating at high temperatures (Ref 232). One of the main challenges regarding the deposition of coatings onto the glass ceramic coatings is the adhesion between the substrate and the coating (Ref 232). Given the brittle nature of the glass ceramic material, grit blasting is not widely used to roughen the glass substrates as the impact of the particles can damage the surface of the substrate. However, it has been found that bilayer coating systems that were deposited on smooth glass ceramic samples did not have sufficient adhesion to the substrate. Although grit blasting can damage the glass ceramic substrate, it can be considered as an efficient manner, especially when accompanied by proper preheating, in providing the adhesion that is required in industrial applications. The micrographs of the coating systems that were deposited on smooth and rough glass ceramic substrates are shown in Figure 11.

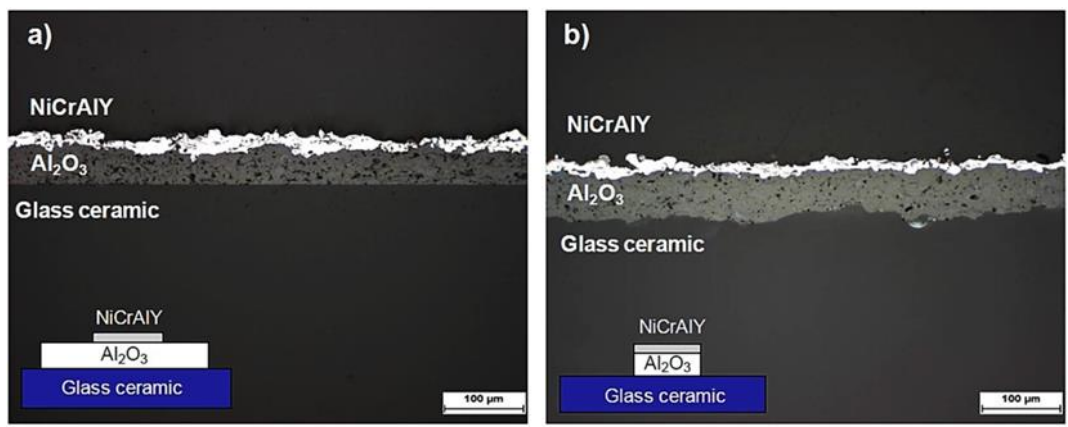

**Figure 11:** *Cross section micrographs of coating systems that are deposited onto a) smooth, and b) grit blasted heated glass ceramic substrates* (Ref 232).

The other main issue associated with direct deposition of metallic coatings on glass or glass ceramic substrates is evolution of critical residual stresses within the coating system due to the noticeable difference between the CTE of the glass substrates and that of the metal coatings that can lead to cracking and delamination at the coating-substrate interface. It was found that the areas that were coated with alumina and NiCrAlY materials had a profound impact on the stability of the bilayer coating system. For some of the samples in which alumina coating was deposited over a wider region, delamination was not observed (see Figure 12(a)). However, for the case in which both layers were deposited onto an accurate geometrical path, delamination of the top coating was observed, which was due to the accumulation of stresses at the edges (see Figure 12(b)) (Ref 233).

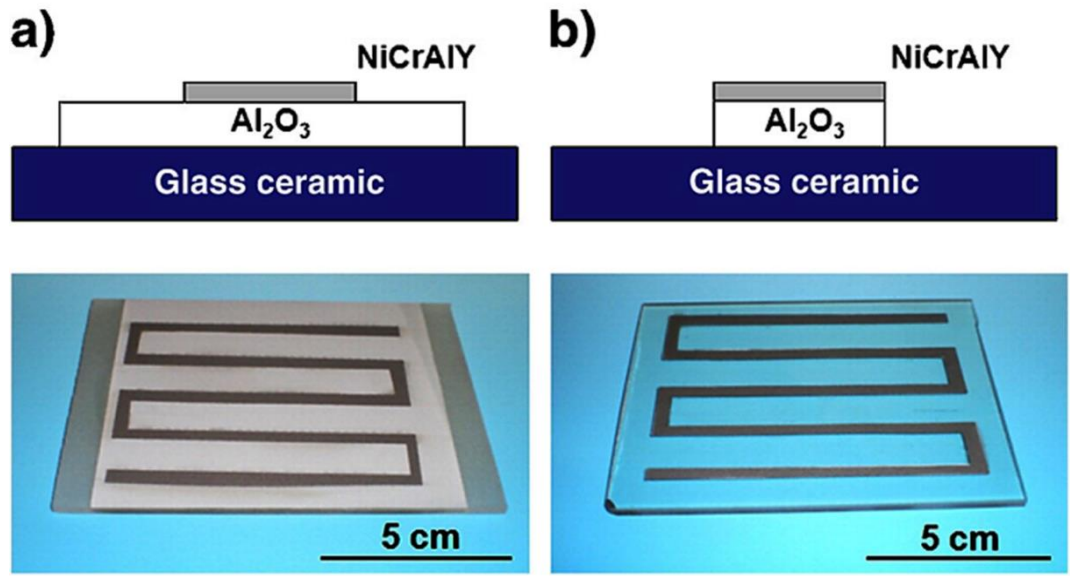

**Figure 12:** *Schematic cross sections and images of a) bilayer coating with wide alumina area, and b) bilayer coating with accurate geometry for both alumina and NiCrAlY coatings* (Ref 233).

Different coating systems, namely bilayer systems that are composed of a ceramic bond coating and a metal top coating, ceramic-metal mixed layers, and ceramic mono-layers were investigated in order to minimize the formation of residual stresses and ensure fulfilment of the required electrical properties (Ref 233). It was found that mono-layer electrically conductive coatings are more stable and have better adhesion to the substrate compared to the conventional bilayer coating systems. From all the developed coating systems in this study, it was found that the best adhesion and stability was achieved for the case in which titania ($TiO_2$) was deposited on the glass ceramic. Furthermore, it was found that addition of NiCrAlY to $TiO_2$ feeding powder can enhance the electrical conductivity of the coating layer. However, delamination of the coating is more probable for cases in which higher content of NiCrAlY is added in the $TiO_2$/NiCrAlY mixed phase. The micrographs that were taken from the cross section of the developed coating systems are shown in Figure 13.

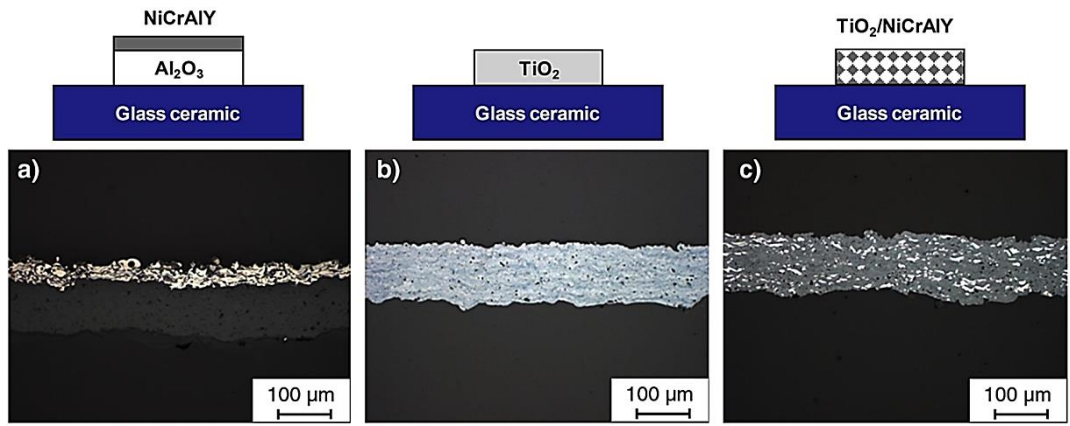

**Figure 13:** *Cross section micrographs taken from a) Al2O3-NiCrAlY bilayer coating, b) mono-layer TiO2 coating, and c) TiO2/NiCrAlY mixed phase coating* (Ref 233).

Tong *et al.* (Ref 181) investigated fabrication of microheaters by micromachining thermal-sprayed 80Ni-20Cr coatings. In this study, thermal spray, as an additive manufacturing process, was combined with micromachining, as a subtractive manufacturing process, to produce functional microheaters. This technique, as described in section 3.4, shows considerable potential in fabricating small scale embedded functional parts within thermal-sprayed coatings. In this study, both combustion and plasma thermal spray techniques were used to deposit the NiCr coating with thickness in the range of 25 to 75 µm on alumina substrate. Ultrafast laser direct-wire technique was used to fabricate patterns based on which uniform heat flux can be produced. Several different patterns for fabrication of the micro heaters are shown in Figure 14.

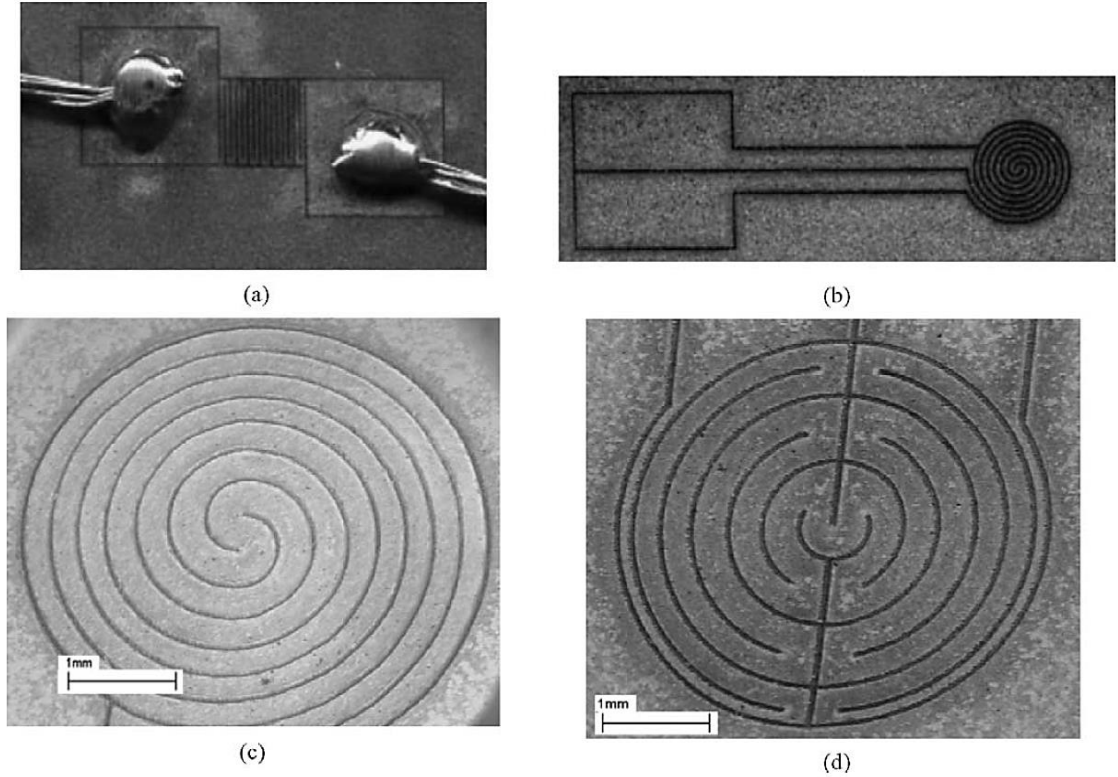

**Figure 14:** *Microheater designs: (a) square-meander, (b) circular-spiral, (c) SEM image of circular-spiral, and (d) SEM image of concentric-circle patterns* (Ref 181).

In this study, cracking and mechanical failure of alumina substrate was observed when the temperature of the heater was increased to 450-500 °C. Therefore, the failure was attributed to the thermal stresses caused by coefficient of thermal expansion mismatch in the multi-layered sample. It was concluded that the bimorph beam bending model can be used to explain the failure caused by thermally induced stresses at elevated temperatures (Ref 181).

Prudenziati *et al.* (Ref 234), developed self-regulated heaters by air plasma spraying process that could operate reliably at temperatures up to 600°C over long periods of time. In this study, the heaters were designed for both planar and cylindrical geometries. Engraved metal masks were used to fabricate the heating elements (Ni, Ni-20Cr, and Ni-5Al) in the shape of meanders as shown in Figure 15.

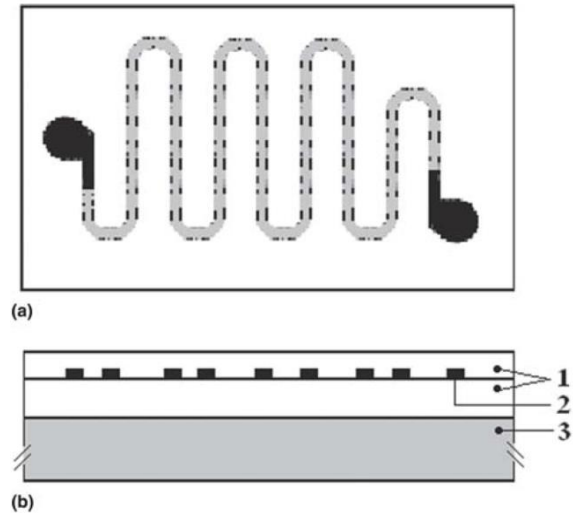

**Figure 15:** *Schematic of the heating plate that consists (1) alumina coatings, (2) heating meander, and (3) metal coupon (substrate) from (a) top view and (b) cross-sectional view* (Ref 234).

Two fabricated heating elements, heating plate and a runner nozzle for a polymer injection moulding apparatus, are shown in Figure 16. Due to the dependence of the heating element electrical resistance on temperature, it was claimed that the metallic heating element could also be used as a smart sensor for direct reading of the heater temperature. Therefore, there would be no need to install a separate temperature sensor. Based on the claimed wide operation temperature range (20-600 °C) and the industrial reliability of the fabricated coating systems, it was concluded that the developed self-regulated heating platforms can be used for applications in the field of high-temperature operating sensors (Ref 234).

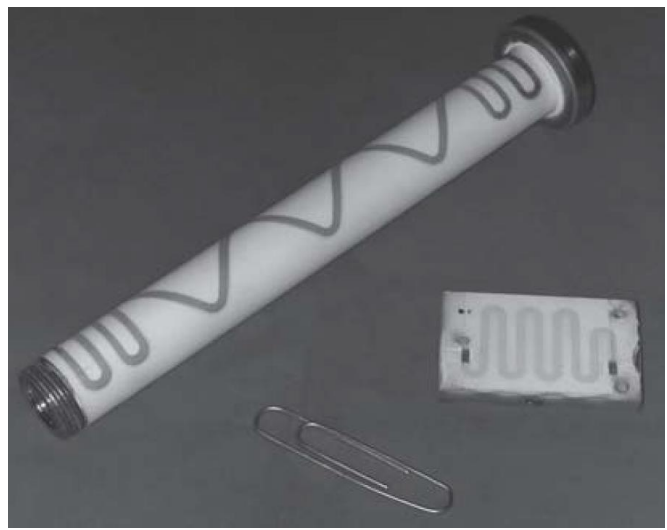

**Figure 16:** *Cylindrical and Planar heaters developed by air plasma spray (APS) process* (Ref 234).

In another study by the same author (Ref 235), the failure of the heating element and the insulating layer was investigated. It was found more serious failure modes were associated with use of NiCr- and NiAl-based heating elements than Ni-based ones. Electrical properties of both thermal-sprayed Ni and Ni20Cr heating elements and the correlation between the microstructure of the coating and the electrical properties of the metal alloys that were deposited by atmospheric plasma spray (APS) and HVOF processes were investigated by Prudenziati and Gualtieri (Ref 236). It was observed that use of fine powders, oxygen-rich atmosphere in the plume, and high temperature flames of APS and HVOF lead to preferential oxidization of Cr and formation of oxide phases (mainly chromite $NiCr_2O_4$ and NiO) that are high resistive materials. The Ni/Cr ratio in the oxidized regions was found to be lower than the value in the surrounding matrix. This indicates that oxidation results in enrichment of Cr in the oxidized areas, which leads to depletion of Cr and enrichment of Ni in the surrounding areas resulting in composition change from 80Ni20Cr to xNiyCr with $x/y > 4$. This forces the electrical transport in the surrounding metal matrix with higher Ni/Cr ratio resulting in higher temperature coefficient of resistance (TCR) values. Furthermore, it was found that the TCR of NiCr that was deposited by APS was considerably less than the HVOF-sprayed NiCr (Ref 236).

Lamarre *et al.* (Ref 230) developed a multi-layered thermal-sprayed cylindrical resistive heater. In this study, alumina dielectric insulator with a thickness of 250-300 µm was deposited onto metallic substrates by plasma spray process. Then, a wire flame-sprayed iron-based alloy (FeCrAl) resistive element with thickness in the range of 75-100 µm was deposited onto the alumina coating with a specific pattern that is shown in Figure 17 by masking technique. The performance and the microstructure of the heating system were analysed after energizing and maintaining the cylindrical heater at a constant temperature of 425°C for up to four months.

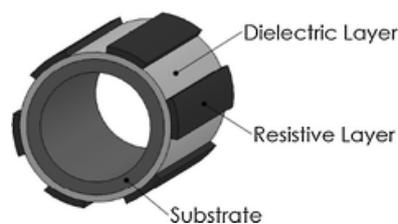

**Figure 17:** *Schematic of a portion of the cylindrical resistive heating system (not to scale)* (Ref 230).

A localized high-stress zone was observed at the intersection point of FeCrAl, alumina, and ambient air, which is highly due to the discontinuity of the applied force on the surface of the alumina coating. Formation of cracks in the alumina layer, which is of great importance in deterioration of alumina layer and failure of the insulator and can lead to short circuiting between the conductive substrate and the heating element, can be observed in Figure 18. Based on the results obtained from the simulation model, it was concluded that reducing the thickness of alumina can decrease the maximum stress, but it increases the average stress in alumina layer.

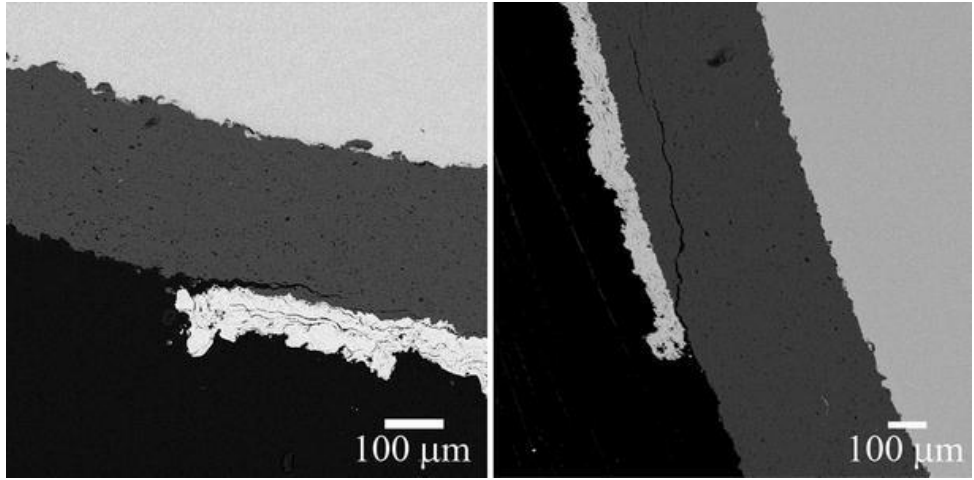

**Figure 18:** *Cross-sectional SEM images taken from the multi-layered heating sample showing formation of cracks at the intersection point after conducting the thermal tests* (Ref 230).

The review of the materials, processes, and reported results with respect to fabrication of multi-layered coatings for heating applications shows that an efficient heating system can be successfully developed by thermal spraying techniques. However, care should be taken to avoid challenges that can lead to malfunction and failure of the coating systems, including the high tensile residual stresses generated during the spraying process, dielectric breakdown mechanism and the resulting short circuiting, and evolution of excessive thermal stress and the consequent delamination of the coating layers, to name a few. Novel approaches were taken recently to fabricate two thermal-sprayed heating systems that can be used for advanced engineering functional applications. Further details on the applications, associated challenges, fabricated coating systems, and the promising results are given in the following sections.

### 3.5.2. De-icing of wind turbine blades

The formation of ice on top of the wind turbine blades and compromise of the surface integrity of the blades can adversely affect the performance of the wind turbines (Ref 237). It has been observed that both mass and aerodynamic imbalance take place even at the initial stages of icing (Ref 238). In some cases, severe icing has resulted in full stoppage of the turbine that has led to significant non-productive downtime and production loss (Ref 237). In harsh sites, the reduction in the annual production has been reported to be in the range of 20-50% (Ref 239). Furthermore, random shedding of ice from the rotating blades can induce out-of-balance loads, which can increase fatigue loading and cause premature mechanical failures in the components of the wind turbine (Ref 240). Aside from the detrimental technical aspects, formation of ice on the blades can pose a safety hazard due to the risks associated with ice throw from the blades (Ref 241).

It is due to the above-mentioned detrimental consequences that utilization of an effective solution that is capable of mitigating the problems associated with accretion of ice seems to be of great importance. In this regard, coatings with anti-icing properties such as ice phobic coatings have been studied extensively; however, it has been proven that the passive anti-icing coatings are inefficient to prevent formation of ice on the blades alone, and their efficiency has been demonstrated only in less severe sites (Ref 242). Therefore, an efficient de-icing method is still needed to remove the ice layer from the blades.

That said, flamed-sprayed coatings as novel heating elements with the potential impact of eliminating the formation of ice on wind turbine blades were proposed by Lopera-Valle and McDonald (Ref 243,244). It was found that flame spraying process could be used to deposit both coating-based heating elements, namely nickel-chromium-aluminium-yttrium (NiCrAlY) and nickel-chromium (NiCr) coatings, on top of the fibre-reinforced polymer composite (FRPC) substrate successfully without any damage to the substrate. In order to prevent damage to the fibres of the composite substrate due to the deposition of high-temperature molten powder particles, a layer of garnet sand was used as an intermediary layer between the FRPC and the metal alloy coating. Furthermore, the garnet sand layer that was formed by sparkling garnet on top of the last layer of epoxy prior to the curing process was also used as the roughening agent. It was found that the use of the garnet sand layer was beneficial for both deposition of the flame-sprayed coatings and also protecting the substrate against thermal damages during the flame spraying deposition process. The NiCrAlY and Ni-20Cr coatings that were fabricated in this research were as thin as 80 ± 15 µm (*n* = 20) and 100 ± 15 µm (*n* = 20), respectively. The SEM images that were taken from the cross-section of the coated FRPC sample are shown in Figure 19(a) and Figure 19(b).

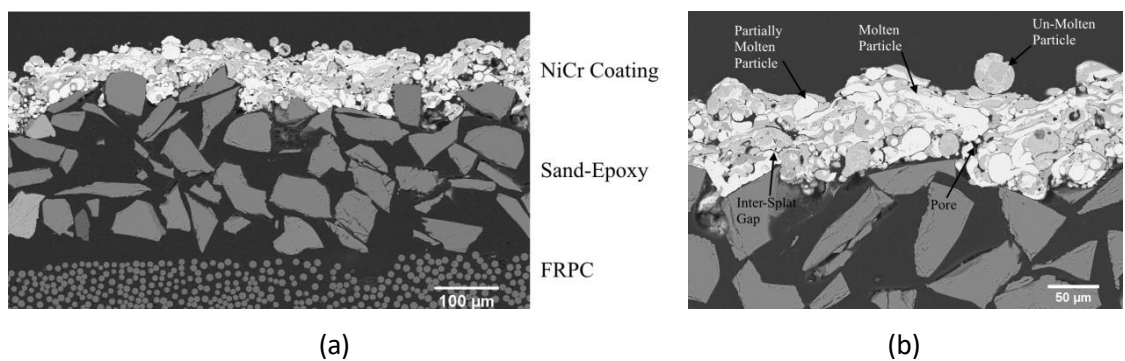

(a)                                                          (b)

**Figure 19:** *a) Low- and b) high-magnification SEM images taken from the cross-section of the coated FRPC* (Ref 243).

It was observed that with a supplied power as low as 2.5 W over 3 V the temperature of the FRPC sample with dimensions of 20 mm × 120 mm was 15°C above the ambient temperature when no air was flowing over the sample (Ref 243). In addition, application of the efficient coating-based heating element was successful in increasing temperature of the FRPC sample under forced convective conditions. In another study by the same authors (Ref 244), it was shown that the embedded coating-based de-icing element was also able to melt the ice that was formed on top of the coated polymer-based composite while the sample was exposed to forced convection conditions.

### 3.5.3. Heat tracing of metallic pipelines used for water conveyance and drainage

It is well-known that the solidification of water inside the pipes that are exposed to temperatures below the freezing point of water for an extended period of time is of great concern. Given the volume expansion of the enclosed freezing liquid upon transformation into ice, continuous formation and axial growth of ice inside the piping systems will result in pressure build-up between the ice plugs and rupture of the pipes (Ref 245). It has been found that in addition to the solid annular ice, the formation of dendritic ice at the very early stages freezing process can also result in blockage of the flow of water (Ref 246). Furthermore, the failure of the frozen pipes is not limited to metallic pipes and it can also occur for the plastic pipes (Ref 247). This undesired phenomenon is of great importance from both financial and safety viewpoints. Financial losses including the damage cost and, especially, plant down time are of great values. The costs due to the damage to the pipes because of the freezing of the enclosed liquid has been reported to be $450 million/year during the 1985 to 1995 period only in the United States (Ref 248). According to another study that was performed by the Insurance Information Institute (I.I.I.), the total financial losses due to the damage from the frozen pipes and the related issues, which is one of major issues that occurs during winters, averaged $1.2 billion annually only in the United States over the last 20 years (Ref 249). In addition to the financial losses, burst of pipes may be found as a hazard to the safety of the workers in the field to the tremendous pressure release upon failure that may result in explosive ruptures.

It is due to the aforementioned serious consequences that developing an efficient heating system that can mitigate the solidification of entrapped water inside pipes is of great importance in industrial sector because installation of the insulation, as the most cost-effective and available solution to reduce the heat loss rate and prolong the duration of freezing (Ref 250), is only sufficient when the stagnant water inside the pipe is exposed to cold environment for a short period of time. In this regard, conventional heat tracing system is used in the industry to overcome this challenge.

Thermal-sprayed coating systems, as electrical resistive heating systems, can be used as an efficient alternative to mitigate the problems associated with frozen pipes. The other advantage that application of the coating-based heating systems offers is the possibility of using them as strain and temperature sensors so that the heating system can act as a closed-loop control system that operates independently. In this respect, any changes in the electrical resistance of the heating element could be monitored. Then, the changes can be attributed to the fluctuations in temperature or potential deformation/damage of the metal alloy coating.

The possibility of deposition of a thermal-sprayed multi-layered coating on a low carbon steel substrate and its functional performance were studied (Ref 251). In this study, nickel-50 wt.% chromium (Ni-50Cr) was selected as the heating element owing to its high electrical resistivity. Due to the electrical conductivity of the steel pipe, an intermediary layer, as an electrical insulation, was deposited between the conductive substrate and the heating element to ensure that the short circuit does not occur and the free electrons do not move through the pipe that has less electrical resistance compared to the coating-based heating element.

Because of the unique electrical and thermal properties of alumina, this material was selected to act as the electrically insulating layer to prevent the short circuit and malfunction of the heating system. While the flow of free electrons from the coating to the steel pipe was obstructed by utilization of alumina due to its dielectric properties, the heat was freely transferred to the pipe and the ice inside the pipe due to the high thermal conductivity of the alumina coating, which is relatively higher than other ceramics. It has been reported by Francl and Kingery (Ref 252) that the thermal conductivity of the bulk alumina is in the range of 29-36 W/mK. However, it is proven that the thermal conductivity of thermal-sprayed coatings is less than that of the bulk dense materials due to the limited bonding at the lamellar interfaces of the coatings (Ref 253).

Cold-sprayed dense copper coating was also deposited onto the NiCr in this study to make an ideal electrical connection between the metal alloy coating and the power supply. The coating-based heating element was heated by way of Joule heating in proportion to the supplied power. Figure 20 shows the multi-layered coating system that was deposited onto the pipe assembly. Furthermore, the installed thermowells and thermocouples that were used for measuring the transient temperature of the ice/water during the heating tests can be seen in Figure 20.

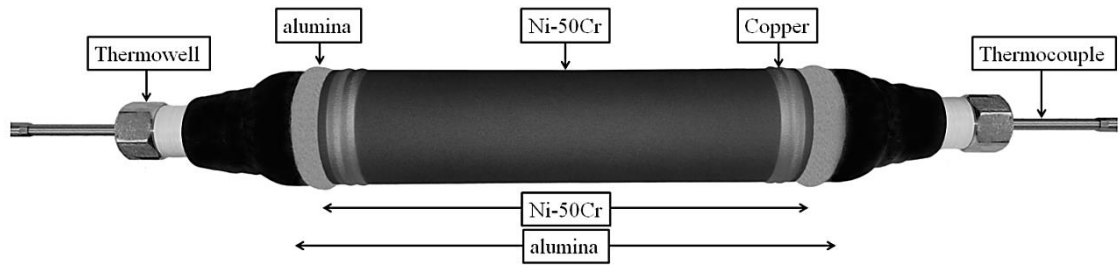

**Figure 20:** *The components of the coated pipe assembly that was used for the Joule heating tests* (Ref 251).

Given the probable damage to the plastic pipes due to the high temperature of the flame and the molten feedstock particles, deposition of the flame-sprayed coatings has been tried only for the case of metallic pipes so far. It was concluded that the combination of the powder materials, namely Ni-50Cr, alumina, and copper, was ideal for the heating task. While the pipe assembly that was filled with solid ice was exposed directly to the circulation of ambient cold air at a temperature of -25°C without any thermal insulation, even a power as low as 20 W was enough to heat and melt the ice within the pipe (Ref 251). For the case in which a proper thermal insulation was wrapped around the pipe, even 10 W was sufficient to heat and melt the ice inside the pipe easily. Furthermore, no electrical or mechanical damages were detected during the microstructural evaluation of the multi-layered coating system by using a scanning electron microscope (SEM). These promising results speak to the high efficiency and operational advantage of the coating-based heating element for being used in industrial applications. Figure 21 shows the SEM images from the coating system after conducting quite a few heating tests with various supplied powers.

It was observed that the measured average thickness of alumina as low as 175 µm (*n* = 10) was enough to protect this layer from dielectric breakdown and the whole coating system from short circuit and the resulting malfunction. Furthermore, the average thickness of the deposited NiCr coating was measured to be as thin as 110 µm (*n* = 10). This shows that only little amount of NiCr powder is required for developing this heating system, and therefore, fabrication of this coating system is feasible from financial viewpoint. Due to the lamellar structure of the coating and the presence of the unmolten particles that is common in flame spraying process, the obtained microstructure of the NiCr coating was porous. The average porosity of the coating was measured to be 7.5 vol.% (*n* = 10). Although presence of pores and microstructural defects in the flame-sprayed coatings can affect the mechanical structure of the coating layers negatively, it can increase the electrical resistance of the coating layer by reducing the effective cross-sectional area of the conductive coating (Ref 254).

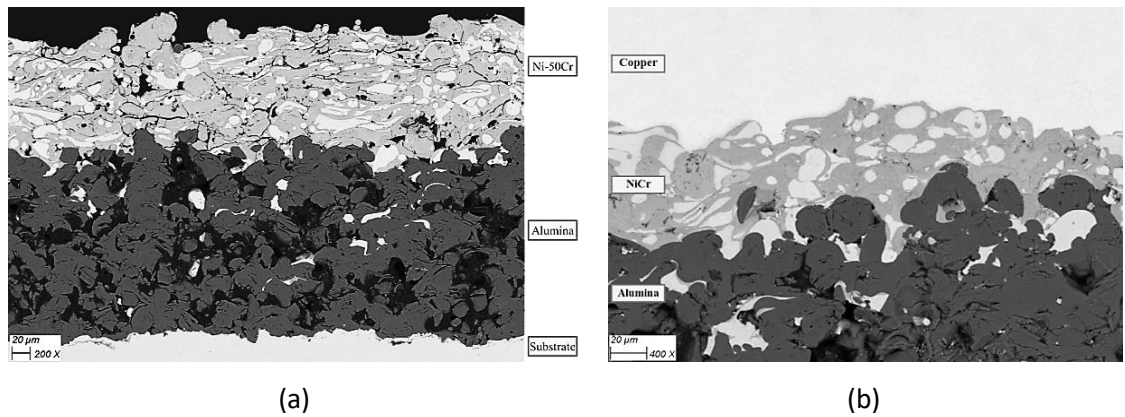

**Figure 21:** *SEM images taken from the cross-section of a) bi-layered coating in the middle and b) tri-layered coating at one of the ends of the pipe assembly* (Ref 251).

Although the deposition of multi-layered thermal-sprayed coatings onto steel pipes for the heating purpose and their functional performance have been successful so far, the durability and service life of such coatings subjected to cycling severe thermal loadings should be assessed prior to adoption of these novel coating systems in the field. The current challenge with the application of these coatings is still the coefficient of thermal expansion mismatch between the brittle alumina ceramic layer and the metal alloy coating and the steel substrate. This can result in evolution of thermal stresses and formation of microcracks in alumina, which would, in turn, deteriorates the dielectric properties of alumina. Application of suspension plasma spraying process that can fabricate coatings with more strain tolerant microstructure or deposition of materials with dielectric properties, but with higher toughness, can be the subjects of further studies in this area in the future.

Furthermore, given the necessity of developing energy efficient heating systems, it is beneficial in industrial applications to initiate the heating of the pipe in advance of the time when the freezing of the enclosed water can cause plastic deformation and damage to the pipe and the coating structures. Given the dependence of the freezing behaviour of the enclosed water on the mechanical properties and wall thickness of the pipe (Ref 255), and the brittle nature of the alumina coating that can only endure very little elastic deformation, it stands to reason the resistive heating system should be operational prior to the end of the first solidification plateau when the volume of the mix of water and ice occupies the whole capacity of the closed system and further transformation of water into solid ice will cause internal pressurization (Ref 255). Although during the early stages of pressurization of the entrapped freezing liquid the pipe undergoes elastic deformation, it is likely that the ceramic coating fails due to the fact that its failure strain is less than the yield strain of the steel pipe. Therefore, better understanding the freezing behaviour of the enclosed water and the potential damage to the coating layers and the pipe for understanding the critical time for initiation of the heating and decreasing the energy consumption of the heating system can also be the subject of future studies. Furthermore, potential application and feasibility of using the metal alloy coatings deposited onto steel pipes as both temperature and strain sensors and the reliability of the obtained measurements by reading and monitoring the slight changes in the electrical resistance of the coating-based heating element could be another subject for further study and research in the future.

## 4. Smart Coatings

Smart coatings are considered the next step in the development of more capable coatings, accounting for their active response to intrinsic or extrinsic events, in contrast to the passive behaviour presented by traditional and functional coatings. In this section, an overview of the most significant developments in the field of smart coatings applied through the use of thermal spraying deposition techniques will be presented.

### 4.1. Sensors

The unique capability of smart coatings to produce an active and coherent response to a wide range of stimuli makes them ideal for the development of novel sensors. The capability to integrate the well-stablished protection and mechanical capabilities of thermal sprayed coatings with a calibrated response to environmental factors, lends itself into sensors for highly demanding environments. The concept was demonstrated on the early 1980s with the registry of a patent for an oxygen sensor based on porous plasma-sprayed zirconia membrane between two electrodes (Ref 256). The capability of the porous membrane to react to the different oxygen partial pressure of the atmosphere was followed by a rapid variation of the electrical resistance between the electrodes. This development opened the path for the use of thermal spray techniques, readily available in the industry and with excellent deposition efficiencies, on specialised applications. To give an idea of the importance of the breakthrough, the same fundamental design is used nowadays to produce gas sensors, although the capabilities have been increased. Gardon *et al.* (Ref 177) report the development and testing of a $NH_3$ APS-deposited sensor based on a mixture of titanium oxides. Similar to the early 1980s concept, the oxygen vacancies on the coating facilitated the adsorption of $O_2$, which transforms into anionic oxygen species, highly reactive towards $NH_3$, producing a change in the electric resistivity. An important point of their work relies on the substrate used to build the sensor on, a thin polymer film, essential for lightweight, flexible sensors. Through careful selection of the deposition parameters, a common, flexible, heat-sensitive material was used to produce a gas sensor using a well-known technique such as APS. Nevertheless, it has been the field of photo-reactive smart coatings that has seen a considerable dedication over the past decades, producing highly capable sensors.

### 4.1.1. Photoluminescence

The application of photoluminescence on smart coatings produces a noncontact technique capable of providing information on environments where access is limited or impossible. The principle is based the emission of photons with specific wavelength related to the electronic structure of the compound. This effect is commonly achieved through the addition of small amounts of fluorescent elements into the coating. The interest on thermal sprayed smart coatings with sensing capabilities is based, for one, in the fact that thermal sprayed coatings have been widely used on high temperature environments, as corrosion resistance and thermal barrier coatings (TBC). This provides a solid foundation base due to their extensively studied properties and behaviours, with the addition of a successful record of industrial implementation. Secondly, accounting for the extreme environments experienced and the significant cost of maintenance shut-downs to perform inspections, a need for externally operated, capable of providing online measurement sensors has been identified. These factors combined has led to the development of systems that, while retaining their thermal and corrosion resistance properties, allow for the measurement of the state of the coating.

As mentioned before, the widely extended use of thermal sprayed coatings as TBC makes this field a suitable choice of the development of smart coatings. As such, considerable work has been carried out aiming to develop reliable thermal barrier sensor coatings (TBSCs). The underlying idea of a TBSC is a ceramic coating doped to a low level (usually below 0.1 wt. %) with a rare-earth (RE) component capable of luminescence, with a few examples on their general structure shown in Figure 22.

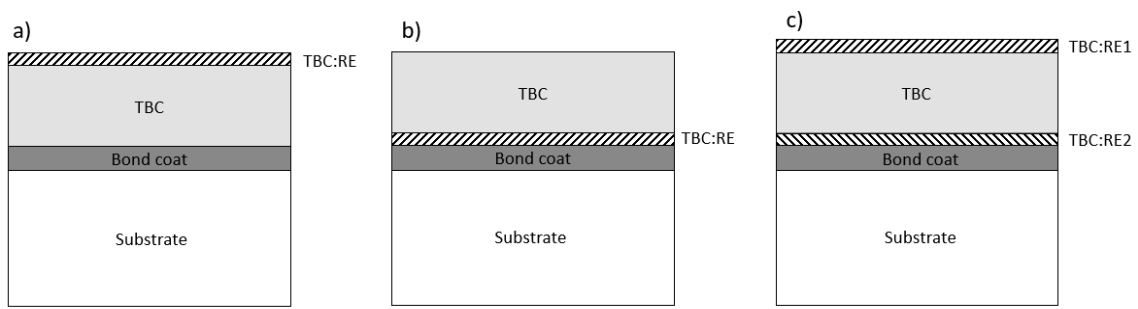

**Figure 22:** *Three different examples of luminescence rare earth (RE) doped layers (striped sections) on thermal barrier sensor coatings for a) temperature measurement at the surface, b) temperature measure within the TBC and c) dual temperature measurement, allowing for heat flux calculations.*

An extensive review published on 1997 by Allison and Gillies (Ref 257) gives a detailed overview of the remote thermometric technique, with the appearance only one year later of an European patent for the invention of smart thermal barrier coatings with sensing capabilities (Ref 258). Two fundamental phenomena are exploited on the development of these sensors. The first, and most common one, is the temperature dependence of the fluorescence decay time (Ref 259–265). This method, which provides reliable, online measurements of the temperature at the surface and in-depth, requires the use of a pulsed excitation source. After each pulse, the emission intensity from the phosphor follows an exponential decay shown on the equation below (Ref 265):

$$I = I_0 e^{-\frac{t}{\tau}}$$

where $I_0$ is the intensity at time zero and $\tau$ is the time constant of said process, which is dependent on the temperature. The second method, with a more limited development, is the measurement of the intensity ratio between two adjacent emitting lines (Ref 266–268). In this case, both a pulsed and a continuous source can be used, as the ratio of two distinct spectral lines from the phosphor is used instead of the emission decay. The intensity ratio of the closely spaced lines on the luminescence process is related to the electron population ratio of the electronic levels, which is temperature sensitive according to Boltzmann equation (Ref 267):

$$\frac{I_1}{I_2} = \frac{n_1}{n_2} = e^{-\left(\Delta E / k_B T\right)}$$

where $I_1$ and $I_2$ are the intensity of the adjacent lines, $n_1$ and $n_2$ are the electron populations of the associated levels, $\Delta E$ is the energy gap between them, $k_B$ is the Boltzmann constant and $T$ is the temperature in Kelvin.

As explained on the work of Gentleman *et al*. (Ref 266), interest on sensing thermal barrier coatings was initially focused on those produced using electron beam physical vapour deposition (EB-PVD), due to the columnar structure associated with this deposition technique, which yields better optical transmission than APS. In their work, they present an excellent theoretical overview of the state and potential limitations of the development of smart sensing TBC, providing at the end a simple, yet effective, system with sensing capabilities as a proof of concept. Due to the wide acceptance of APS in the industry, better performances have been pursued on thermal barrier sensing coatings deposited using this technique. The two main concerns when developing TBSCs are the introduction of an exogenous fluorescent component without altering the physical properties and durability of the coating, and the deposition of a coating capable of transmitting the optical signal from the embedded sensing layers. Early studies proved the suitability of rare earths such as Eu and Dy as dopants, not affecting the durability of the coatings on cyclic thermal testing. Regarding the burying depth of the sensing layer, continuous efforts are being made to approach the performance of such sensors to their theoretical capabilities.

The first example of a APS-produced sensing TBC is reported by Eldridge *et al*. (Ref 259), on the commonly used YSZ as thermal barrier coating, and a 25 µm thick $Y_2O_3$:Eu as the dopant layer, being applied at the surface. These methods circumvents the known problems presented by pyrometers, which can also measure the radiation from the hot environment, or thermocouples, which can yield inaccurate results at high temperatures (Ref 267). Although online temperatures up to 1100 °C were measured from a depth of 100 µm into the coating, experimental work was carried out on a free-standing coating, which cannot properly simulate the conditions faced by TBCs during service on industrial components. More appropriate conditions were simulated by Chen *et al.* (Ref 261) with Ni-based superalloy substrates, a NiCrAlY bond coat, a 50 µm thick dopant layer of YSZ:Dy and a 550 µm commercial YSZ overcoat, such as the schematic shown on Figure 22(b). The sample was deposited using a production standard APS process. A temperature sensing range of up to almost 1000 °C was achieved, further proving the use of the fluorescence decay time. In addition to the analysis of the sensing capabilities of the dopant layer, their work studied the potential effect of the addition of this layer to the thermal durability of the coating. Although delamination occurred at the TBC/bond coat interface during thermal cycling, the authors classified this failure to be normal of a conventional TBC, with no detrimental effects noted due to the dopant layer, proving a good chemical compatibility in the system. A similar system was studied by Feist *et al.* (Ref 263), using as well a NiCrAlY bond coat and a 50 µm thick YSZ:Dy on top of Haynes 230. In this study, a more thorough structure was investigated, with a commercial YSZ overcoat with thicknesses varying from 50 to 500 µm on top of the dopant layer. Their work confirmed the similarity in microstructure of standard TBCs, and therefore no stability issue were to be expected, along with a calibrated range of measured temperature up to 800 °C for the 500 µm thick YSZ overcoat. Additionally, theoretical studies of the system indicated that the multi-layered structured could be used to measure coatings with a maximum thickness of 1.33 mm, beyond the values of conventional TBCs, providing a theoretical backup for the use of TBCS. Further investigations on the potential of TBSCs were published by Heyes *et* *al.* (Ref 265) reporting online temperature measurements, via the decay time, on YSZ coating with a 50 μm thick YSZ:Dy dopant layer, achieving a sensing depth of up to 500 μm with an upper temperature limit of 800 °C. Even though the sensing limit of 800 °C still leaves room for improvement, the use of a standard APS deposition technique, the fact that few TBC reach thickness close to 500 μm and their extensive investigation of the durability of the TBSC in comparison with regular YSZ TBC, makes their work a clear example of the potential application of luminescence temperature sensors on thermal sprayed coatings within the current industrial standards. Precisely, a wider temperature range is reported by Abou Nada *et al.* (Ref 260) on the development of both surface and 300 μm deep, online measurements. The use of a sandwich structure with a thin DySZ layer at the surface and $Y_2O_3$:Eu at the bond coat/TBC interface with a structure shown on Figure 22(c), both with a thickness of approximately 10 μm, is proven. The temperature range for the $Y_2O_3$:Eu was set to be between 400 – 900 °C, whereas with the use of the DySZ topcoat the upper limit was extended up to 1000 °C. In addition, the system was tested on a Siemens burner and field-tested to service conditions, proving its commercial value.

Another interesting application of luminescence on the field of temperature sensing is the use of irreversible changes on the optical properties of the phosphors to determine the offline, or thermal history of a component (Ref 264). However, limited use of thermal spraying techniques has been reported, in favour of deposition techniques which allows the use of thermal paints, such as screen printing or sol-gel (Ref 262,268).

The remote measurement of components temperature is not the only application that TBSCs have. The flexibility provided by the use of luminescence-doped layers allows for a wide range of devices, as illustrated on Figure 23 and Figure 24. The schematic of a "red-line" sensor, capable of signalling the critical erosion on a coating due to the exposure of an underlying doped layer when the top layer is eroded away, is shown on Figure 23.

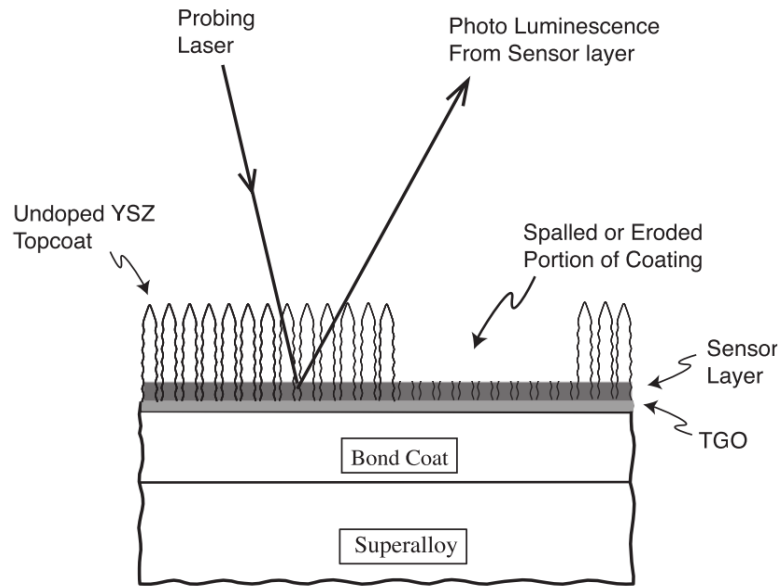

**Figure 23:** *Basic representation of a thin, doped layer at the YSZ-bond coat interface, acting as a "red-line" sensor, indicating when a specific level of erosion has been reached* (Ref 266).

Another potential application, the so-called "rainbow" sensor, with both erosion and temperature profiling capabilities, is presented on Figure 24. The concept is based on a TBC with several layers doped with different luminescence elements, each one of them producing a signal with a distinguishable and characteristic frequency under laser illumination. As the top layers disappear due to the effect of erosion, their corresponding peaks cease to show in the corresponding spectra, marking the progression of erosion.

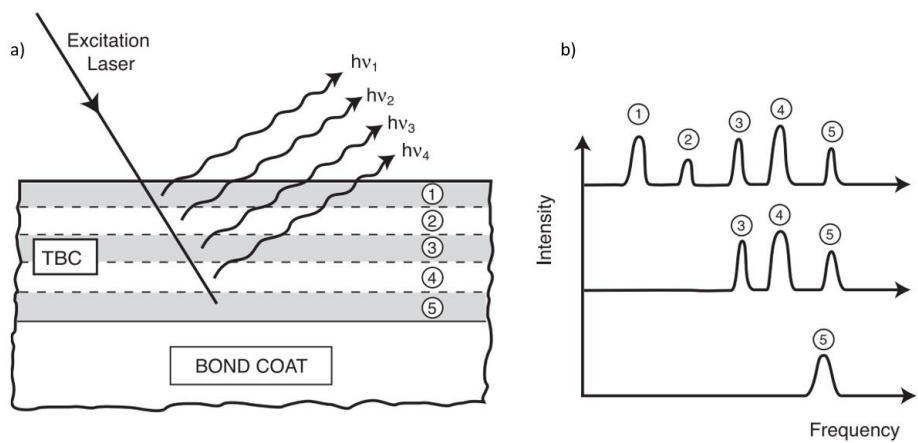

**Figure 24:** *a) "Rainbow" sensor schematic structure with multiple layers doped with different luminescence materials, providing profiling capabilities due to the distinguishable associated peaks of each layer, as shown in b) where the top layers show no presence of luminescence peaks due to erosion* (Ref 266).

A "rainbow" type structure was proposed by Gentleman *et al.* (Ref 266) and as early of 1989 on a registered patent by Amano *et al.* (Ref 269) on a thermal barrier coating. However, as in the case of thermal history sensors, preference to non-thermal spraying techniques has been given (Ref 270,271). In this case, EB-PVD represents a better candidate for its previously mentioned improved optical transmission when compared to standard APS. Still, erosion of TBSC requires additional investigations to provide a reliable measurement of the status of the coating. The technique however shows potential, as shown by the successful development of temperature TBSCs.

### 4.1.2. Embedded sensors

The "smart" behaviour of a coating can also be achieved through the embedment of a stand-alone sensor within the thermal sprayed layers, effectively emulating the effect described on previous sensing smart coatings. This provides a coherent and calibrated response to external stimuli. The concept is based on the integration of an externally manufactured sensor, such as the strain gauges or thermocouples mentioned on section 3.4, into a multi-layered coating, with a structure schematically shown on Figure 25.

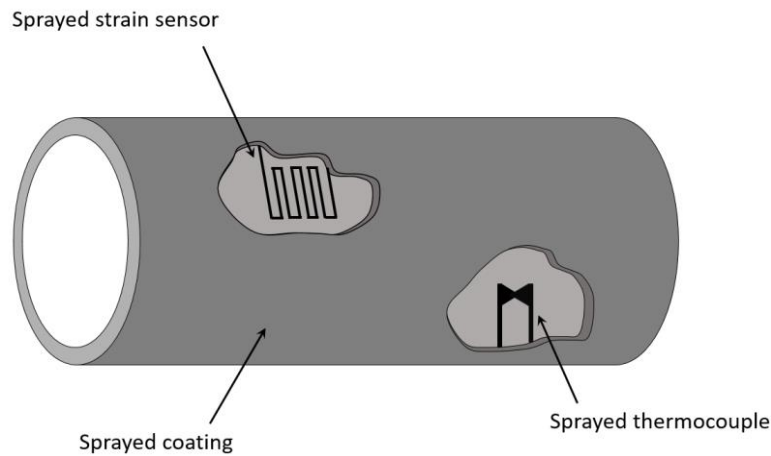

**Figure 25:** *Schematic representation of two embedded sensors within a cylindrical coated piece. Redrawn from* (Ref 272).

Embedded sensors present several advantages over intrinsic smart sensing coatings, since sensor and coating can be designed separately and then integrated. The only requirement is that the embedded sensor does not chemically interact with the coating, or excessively modify its physical and mechanical properties. The concept was first described by Fasching *et al.* (Ref 272) in 1995, including the description of two such embedded sensors, a sprayed thermocouple and a humidity sensor. Although the reported sensors presented faults and needed further optimisation, they represented the proof-of-concept for embedded sensors using thermal spray techniques that promote further research.

Considerable work has been carried out at the State University of New York, Stony Brook, NY in the development of more precise and streamlined embedded sensors. Thermocouples, as mentioned before, have been ideal platforms for embedded sensors produced via thermal spraying, due to their morphology and material requirements. An examples of such sensor is presented by Longtin *et al.* (Ref 273) using 62Cu/38Ni and 80Ni/20Cr to produce an E-type (constantan 60Cu/40Ni and chromel 90Ni/10Cr) thermocouple on top of a thermal sprayed alumina layer. An additional layer of alumina was deposited on top of the thermocouple to prove its embedded capabilities. The sensor was tested at temperature ranging from 200 °C to 800 °C with a commercial E-type thermocouple to verify its performance, proving excellent linearity and repeatability. A K-type (chomel and alumel 95Ni/3Mn/2Al/1Si) thermocouple was produced using 80Ni/20Cr and 95Ni/5Al for its use at higher temperature, up to 1200 °C, demonstrating once more an excellent performance compared with commercial thermocouples. The basis for a strain gauge is also reported, although due to constrain in the minimum feature size achievable with the deposition technique, in the range of 200 µm, no results are reported.

A commercial-ready set up for embedded temperature and heat flux sensors is reported by Gutleber *et al.* (Ref 274) in collaboration with MesoScribe Technologies (Ref 275) using specifically designed plasma spraying process with dynamical aperture control to achieve the required linewidth. The system was formed by a Ni-based superalloy substrate (Hastelloy X) with MCrAlY type bondcoat and YSZ as the TBC overcoat. The effectiveness of the sensors was tested on bare (no coating on top of them) and embedded state, producing K-type thermocouples and heat flux sensors. Both sensors were tested to temperatures up to 800 °C in comparison with commercial sensors, showing excellent agreement. Durability tests were also performed on a cyclic burner rig with varying temperatures between 75 °C and 1050 °C to demonstrate the commercial applicability of the sensor under service conditions. The commercial application of an embedded resistive strain sensor (Ref 276) produced using thermal spraying and laser machining methods has also been reported, although specific information of materials deposited and geometry of the patterns has not been disclosed. This device showed promising results in terms of sensibility and accuracy, although further research into the impact of real service conditions into the sensing capabilities of the system would be required to assess its applicability into the industry.

### 4.2. Self-healing

The appearance of cracks and other defects within coatings, sometimes being too small or buried deep within the material to be detected, is a recurring problem that affects the coating mechanical properties and its correct functionality. The motivation for coatings that can repair themselves without any external intervention is clear, although its realisation has not been possible until recently. The concept of self-healing within a coating was first demonstrated by White *et al*. (Ref 277) in 2001. Their work, schematically shown on Figure 26, achieved the autonomic healing of a polymer composite. The composite was made of a mixture of a commercial epoxy resin with diethylenetriamine (DETA) via the microencapsulation of a healing agent, dicyclopentadiene (DCPD). The healing agent was released due to crack propagation within the matrix, reacting with the catalyst embedded in the matrix and producing the polymerisation of DCPD. This would create a tough polymer network that allowed the recovery of up to 75% of the initial toughness of the coating.

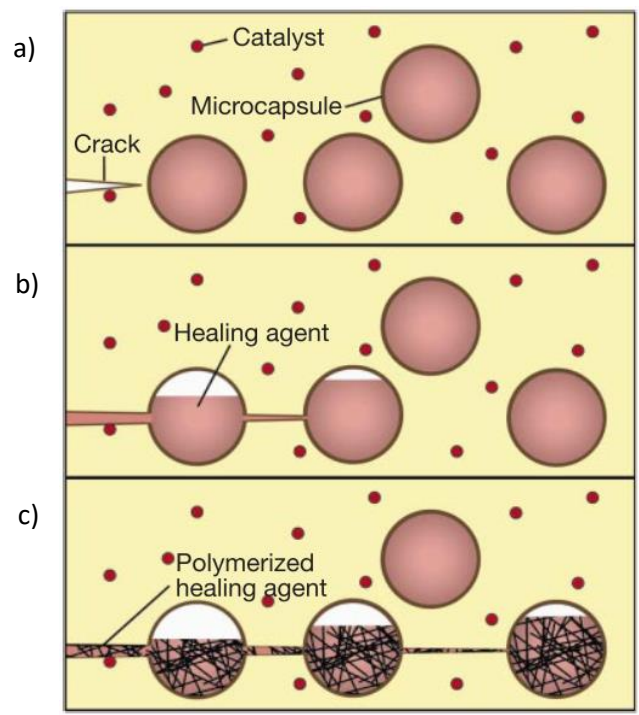

**Figure 26:** *Schematic representation of the self-healing mechanism reported. a) The process is initiated when a crack in the matrix reaches a microcapsule, b) releasing the healing agent, which reacts with the catalyst embedded within the matrix, and c) the polymerisation takes place, effectively closing the crack.* (Ref 277).

Their work can be seen as the proof-of-concept that allowed the development of novel self-healing coatings, due to the flexibility of the system, with multiple possibilities for coatings compositions, healing agents and trigger mechanisms. One such example involves the detection of pH changes, caused by the presence of redox activity, indicative of corrosion on the aluminium substrate and therefore the presence of a fail in the protective coating, as the trigger mechanism for the self-healing effect (Ref 278). The addition of calcium carbonate microcapsules loaded with corrosion inhibitors such as cerium ions and salicylaldoxime, which dissolve if the local pH is decreased, into a commercial water-based epoxy paint, allowed the creation of a smart coating with self-healing capabilities, effectively hindering on-going corrosion in the Al substrate. A different development was reported based on the use of moisture as both the trigger mechanism and the catalyst for the self-healing reaction (Ref 279). Silyl ester was micro-encapsulated and added to different commercial organic epoxy coatings, which were applied to aluminium substrates. After scratches were produced, environmental moisture reached the microcapsules which reacted forming a polymeric hydrophobic layer, further densifying on continuous exposure to the environment, accomplishing both the protection of the scratch, reducing its wettability to corrosive agents, and the healing of the coating.

Polymers represent the prime example and obvious starting point for self-healing materials, due to the ability to repair damage through cross-linking reactions. However, the application of this strategy to purely metallic components, traditional example of thermal sprayed materials, requires a different approach proving a difficult challenge that has hindered the development of an effective solution. For instance, volume increase of phase transformations in metallic coatings, which can effectively block the cracks within the coating, has been proposed. An example of this advantageous phase transformation is exploited by Grünwald *et al*. (Ref 280), in which APS is used to deposit $Mn_{1.0}Co_{1.9}Fe_{0.1}O_4$ (MCF) coating on top of a chromium containing steel, simulating the interconnectors in solid oxygen fuel cells (SOFC). Such components are known to suffer from chromium evaporation at the operating high temperatures, as shown in Figure 27(a), which leads to the degradation of the SOFC, leading to the need of a dense, stable layer which prevents the gaseous chromium oxides from escaping. MCF thermal sprayed coatings were found to present self-healing capabilities in account of the mentioned phase transformation during heating at 500 °C in air, which promoted the oxidation of the internal surfaces of open porosity, exposed to higher oxygen partial pressures as seen in Figure 27(b), presenting a theoretical volume expansion of up to 20%. The formation of the oxide layer in the inner surfaces, Figure 27(c), gradually blocks the open porosity due to the elevated volume change of the phase transformation, which eventually leads to the complete closure of the channels, Figure 27(d), and the subsequent reduction in oxygen partial pressure. From this point, oxygen ingress is limited to solid state diffusion, much slower than gas phase diffusion, considerably reducing the formation of additional oxide. A measurement of the effect of the self-healing process is represented by the decrease of porosity from 12.4% on as-sprayed condition to 2.4% after annealing at 500 °C for 3 h.

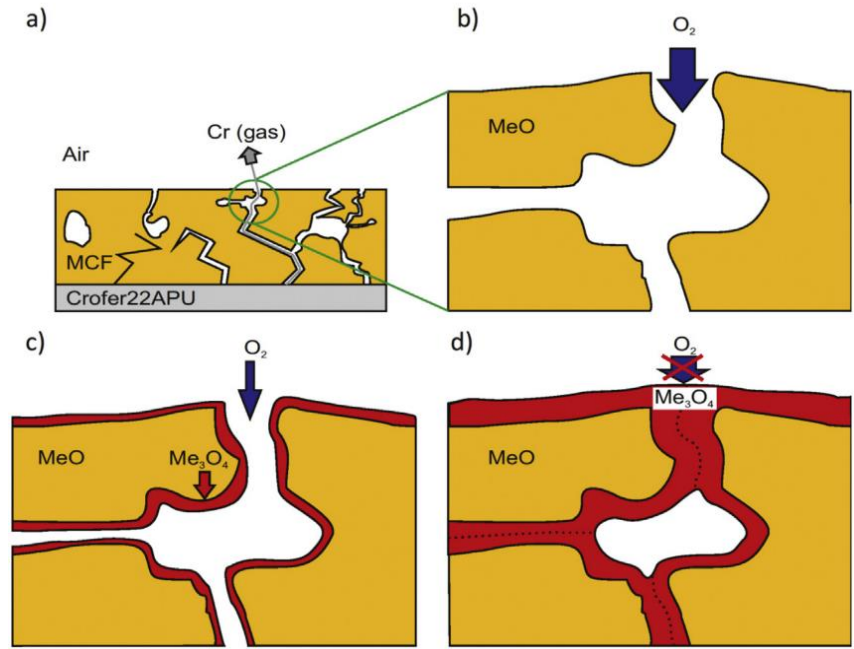

**Figure 27:** *Schematic process of self-healing through phase transformation. a) Cracked MCF coating which allows the evaporation of gaseous chromium oxides from the chromium containing steel substrate and b) detail of the $O_2$ ingress through the cracks. c) Phase transformation into the oxide due to the presence of oxygen in the internal surfaces which leads to d) the closure of the cracks and self-healing of the coating* (Ref 280).

Nevertheless, as it was reported by Vaßen *et al.* (Ref 281), the high levels of stress expected from such a significant volume expansion led the tested samples to experience noticeable bending and large pores appeared close to the coating/substrate interface, which represent risks for the structural integrity of the coating, prone to delaminate under such conditions.

A similar pathway was followed by Derelioglu *et al.* (Ref 282) with the addition of boron alloyed $MoSi_2$ into APS-sprayed yttria stabilized zirconia. The presence of $MoSi_2$ (B) particles acts as healing points, oxidizing once exposed to oxygen. To limit the activation of the healing process until encountering with a crack, the particles had to be protected with a thin, oxygen impermeable layer, such as alumina (Ref 283). Once the propagation of a crack within the coating encounters an encapsulated $MoSi_2$ (B) particle, amorphous silica forms, flowing along the crack. $SiO_2$ further reacts with the YSZ matrix, causing the appearance of $ZrSiO_4$, which acts as self-healing material. Although the development of self-healing TBCs would be highly beneficial, further studies on the impact of the addition of $MoSi_2$ (B) to the thermal conductivity of YSZ are required to assess the potential benefit of the healing effect. This line of work was extensively studied under the SAMBA (Self-Healing Thermal Barrier Coatings) European project, using spark plasma sintering on YSZ (Ref 284) and YSPZ (Ref 285), which reported theoretical simulations and experimental measurement of the effective thermal conductivity of YSZ with different volume fractions of $MoSi_2$ (B) particles. Their work concluded that an addition of 13.5 vol. % increases the effective thermal conductivity from an overall value of 1.5 to 2.5 $Wm^{-1}K^{-1}$, while the presence of 21.8 vol. % produced an increase up to 4.5 - 5 $Wm^{-1}K^{-1}$, effectively rendering the coating inappropriate for thermal barrier applications.

As summarised in this section, smart self-healing coatings is an active field of research, with many potential applications. Although at the present time the extensive investigation of polymers as self-healing coating material and their enhanced self-healing capabilities accounting for the flexibility in the chemistry available, limits the introduction of thermal spraying techniques, promising lines of work have been outlined in the field of metallic and ceramic smart self-healing coatings.

### 4.3. Self-lubrication

The development of coatings which are capable of maintaining a low friction coefficient while withstanding high loads has allowed mechanical components affected by wear damage to increase their service life and efficiency. $MoS_2$ and graphite have been a common choice as solid-lubricant coatings due to their low friction and the capability to produce a low friction transfer film even in dry conditions (Ref 286). Nevertheless, the deposition technique plays a critical role in the hardness of the coating, which can lead to soft coatings (Ref 287,288). Thermal sprayed coatings have been successfully applied in high-wear environments proving excellent tribological properties, making them an ideal candidate for the integration of self-lubrication. For instance, the application of self-lubricating coatings for engine components found in many industries, such as aerospace or power generation, where high strength, corrosion and heat resistances are required on top of their functional behaviour.

In order to achieve this, Li *et al.* (Ref 289) studied the high temperature stability and tribological properties of NiAl-Mo-Ag and NiAl-$Cr_2O_3$-Mo-Ag composite coatings, with a 10 wt.% of Mo and 10, 15 or 20 wt. % of Ag, deposited via APS up to 900 °C. The use of NiAl as the matrix, extensively used on high temperature application in account for its excellent oxidation and corrosion resistances, with the addition of compounds which mitigates its high-temperature wear rate (Ref 290,291), constitutes an ideal candidate for a promising high temperature self-lubricated thermal sprayed coating. Ag-Mo has been reported to have excellent lubricating performance in a wide range of temperatures (Ref 292,293), while $Cr_2O_3$ increases the hardness and wear resistance of the matrix (Ref 294), making them both a suitable addition to the NiAl matrix. The results showed that friction coefficient and wear rates were reduced up to 15 % with increased content of Ag across the whole temperature range studied. The authors suggest the formation of a silver molybdate compound lubrication film as a result of the interaction between Ag and $MoS_2$ formed during the wear test (Ref 295). The addition of $Cr_2O_3$ initially decreased the adherence of the coating, which recovered, and even enhanced this property after annealing at 500 °C. The annealed NiAl-$Cr_2O_3$-Mo-Ag coating showed an slight improvement on the friction coefficient, and a noticeable decrease of the wear rate when compared to the best performing NiAl-Mo-Ag tested, with almost 3 times lower wear rate across the entire temperature range. Although these results show promising self-lubricating capabilities, the need for a heat treatment raises the question of whether additional exposure to high temperatures could affect the phase composition, and therefore the adherence and tribological properties.

Nevertheless, the development of smart self-lubricating coatings, aiming to withstand longer periods of operation without shutdowns or maintenance, independently maintaining their tribological properties to the required level, has attracted attention. The most successful attempt has been made incorporating microcapsules filled with various liquid lubricants, aiming to provide a reservoir available when wear and erosion, potentially caused by the absence of other sources of lubrication, are increased due to absence of lubricant. This approach was initiated by Armada *et al.* (Ref 296) and Espallargas *et al.* (Ref 297), investigating the microencapsulation of liquid lubricants into thermal sprayed coatings, mimicking the system developed by White *et al.* (Ref 277) and covered in the self-healing section. A flame-sprayed Nylon coating, with the addition of lubricant-filled (commercially available polyalphaolefin (PAO) and silicone oil were used) polyurea (PU) microcapsules, was deposited onto carbon steel. To achieve the integration of the Nylon coating with the microcapsules, a hybrid set-up was used, based on a traditional flame spray process for the Nylon powder and an external injector/atomizer. Further details on the hybrid deposition set up are described elsewhere (Ref 298). Upon rupture due to wear, the microcapsules release the lubricant, as it can be seen on Figure 28, reducing the friction coefficient as much as 3 times when compared to nylon coating. This allows for longer service periods with optimal performance, even if external lubricant supply to the component is low, which present an industrial advantage, as reflected by the patent registered on this system by the same authors (Ref 299).

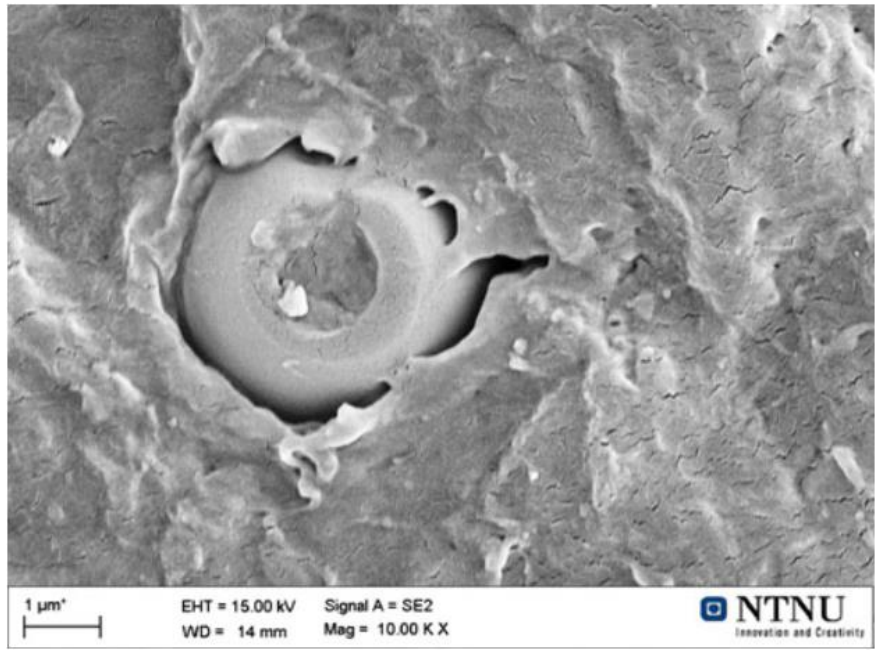

**Figure 28:** *Secondary electron image of a microcapsule after the wear test. The wear and friction caused the microcapsule to break, releasing the lubricant inside. In the image the empty space inside has been then filled with debris* (Ref 296).

The flexibility of matrixes that can be deposited and liquids to be microencapsulated provide a wide range of options, as demonstrated on a later publication (Ref 300), where three different thermal spraying techniques (wire electric arc, atmospheric plasma and high velocity oxy-air fuel) were used to deposit FeNiCr alloy onto carbon steel with added PAO-filled PU microcapsules. The lubricating properties of the different coatings were investigated with respect to the different deposition techniques used and the microcapsules size. Their results show that arc spray produced the coatings with higher porosity, capable of better accommodating the microcapsules, which showed the best self-lubrication performance, without any noticeable difference regarding microcapsule size. The lower temperature and particle velocity of the technique also plays a fundamental role, ensuring that the microcapsules do not suffer damages during deposition.

The development of additional ceramic and metallic matrixes with tailored microcapsules to the thermal spray deposition techniques chosen, and filled with liquid for different applications, such as corrosion inhibitors or self-healing components, would represent the next step in smart thermal sprayed coatings, with a vast potential in several industries were these coatings are already widely applied.

5. **Conclusion**

The introduction of functional coatings expanded the capabilities of thermal sprayed coatings by adding a novel functionality on top of the passive protective characteristics already present. These new capabilities present unique opportunities, such as anti-microbial coatings on equipment and appliances on sterile environments, limiting the surface contamination and cross-transfer of pathogens. On the other hand, well-known thermal sprayed coatings, with a proven record of success on surgical implants can be improved by adding anti-microbial agents that reduce rejection after surgery. The new functionalities would benefit other areas such as the protection of ships hulls and marine equipment against the attachment and growth of foulers, with a severe impact on efficiency and economic turnover.

With the precise aim to provide an increased level of protection, functional coatings that effectively reduce the impact of harmful phenomena caused by environmental conditions have been developed. A clear example would be the introduction of hydrophobic capabilities, presenting a dual purpose. Firstly, water droplets rolling over a hydrophobic surface drag dirt, providing a self-cleaning effect. Secondly, a reduced contact time between the liquid environment and the surface reduces the impact of corroding elements present in the environment. Finally, an important development has been the use of the conducting capabilities of deposited materials. Once more, this characteristic has a dual application, as the deposition of electrical or magnetic materials on a tailored pattern represents a huge opportunity in the field of micro-electronics. Nevertheless, another implication is the creation of resistive heating elements, which provides an invaluable tool for the de-icing of components exposed to harsh environments, such as wind turbine blades, or the prevention of the solidification of water on pipelines.

Despite the great advancements that functional coatings have represented, their applications are limited by an inherent passive behaviour on their interaction with their environment. Therefore, the next step in the development of more capable coatings relied on the introduction of active capabilities with a coherent respond to stimuli, or smart coatings. From the definition itself, the first natural application of smart coatings would be as sensors. Particularly successful has been the application of the photoluminescence effect on rare earth-doped thermal sprayed coatings on high temperature environments, which provides a remote, live or historic measurement on harsh environments. Following this line of thought, sensing capabilities have been achieved including embedded sensors. Although the concept is fairly simple, a traditional sensor is integrated within a component by being placed in between consecutive deposited layers of the coating, its fabrication presents unique challenges such as proper thermal expansion matching, low impact on the finished shape of the component and same service life-time as regular components. Beyond its application as sensors, smart coatings have been successfully developed with another active capabilities, being self-healing coatings a prime example. Within the first variant, traveling cracks encounter microcapsules filled with the healing agent, being released and triggering a reaction in the matrix that would partially restore the properties of the damaged coating. Although the results are promising, coatings require a tailored composition and proper deposition techniques to allow the embedment of microcapsules with the self-healing agent, hindering its application as mass-produced components. Alternatively, self-healing has been proven through the use of phase transformations triggered by the presence of cracks. Such phase transformation implies a volume expansion that effectively seals and restores the cracks. Nevertheless, the concept requires delicate tailoring to ensure that the phase transformation, and associated volume expansion, does not reach critical levels that would cause excessive stress within the coating. Inspired by the microencapsulation of self-healing agents, a new line of work has been developed for the production of self-lubrication coatings. Parting from the solid lubricant coatings already present in thermal spray, self-lubrication makes use of a reservoir of lubricant liquid encapsulated within the coatings itself, achieving a controlled release as the wear reveals and damages the microcapsules. This process ensures the presence of lubricant for longer periods of time, and despite the fact that its effect is limited in time as the reservoir will eventually be depleted, it could present a fail-safe mechanism ensuring proper lubrication of critical components.

In conclusion, functional and smart coatings represent the crystallisation of the wide range of capabilities that thermal spray offers. The successful track record presented in the past, along with the new opportunities provided by the latest developments, points out to an exciting future. Further research will allow to comprehend the deposition and bonding mechanism on multi-component coatings, key to the addition of functional and smart abilities. In addition, the impact that these modified compositions have on the performance of the coatings needs to be thoroughly understood to ensure that the same level, if not improved, of performance is achieved. With these goals in mind, the next steps required for functional and smart coatings would be, first, to translate the current knowledge acquired into readily available feedstock materials tailored to the chosen thermal deposition technique and desired functionality. The great strength of thermal spraying, its flexibility, represents a disadvantage when trying to accommodate all the requirement of the different deposition techniques. The new developments will have to leave the "proof-of-concept stage" and present a sound production process compatible with the standards of the industry, if possible, with minimal deviation from their current deposition techniques. Secondly, this broadening of the current developments aiming to reach penetration into the industry needs to be accompanied by further research into the foundations of functional and smart coatings. The current state of the field proves that there is a vast potential, but some key concepts are still not fully understood. A deeper knowledge of the relationship between deposition parameters and final performance, with particular emphasis on the microstructural causes, will allow to increase the performance efficiency of functional and smart coatings avoiding losses due to unwanted features or undesired phase transformations.